\documentclass[12pt,a4paper]{article}
\usepackage[latin1]{inputenc}
\usepackage[T1]{fontenc}
\usepackage[psamsfonts]{amsfonts}
\usepackage{amsmath,amssymb,mathbbol,latexsym,stmaryrd}
\usepackage[dvips]{hyperref}
\usepackage{calc}
\usepackage{a4wide}
\usepackage[all]{xy}
\begin{document}

\newcommand{\lfiber}[4][]{\ensuremath{\xymatrix@1{ {#2} \ar[r] & {#3} \ar[r]^-{#1} & {#4} }}}
\newcommand{\fiber}[3]{\ensuremath{ {#2}( {#3},{#1}) }}
\newcommand{\tq}{\tilde{q}}
\newcommand{\tQ}{\tilde{Q}}
\newcommand{\tE}{\tilde{E}}
\newcommand{\El}{E_{\lambda}}
\newcommand{\pso}{\Psi^{*} \omega}
\newcommand{\tom}{\tilde{\omega}}
\newcommand{\hr}{\hat{r}}
\newcommand{\hn}{\hat{n}}
\newcommand{\hd}{\hat{\mathrm{d}}}
\renewcommand{\d}{\mathrm{d}}
\renewcommand{\L}{\mathrm{L}}
\newcommand{\R}{\mathrm{R}}
\newcommand{\cotan}{\:\mbox{\rm cotan}\:} 
\newcommand{\ttheta}{\tilde{\theta}}
\newcommand{\tT}{\tilde{T}}

\newcommand{\zo}{{\cal Z}_0}
\newcommand{\ck}{{\cal K}}
\newcommand{\cL}{{\cal L}}
\newcommand{\cM}{{\cal M}}
\newcommand{\so}{{\cal S}_0}
\newcommand{\cK}{{\cal K}}
\newcommand{\wo}{{\cal W}_0}
\newcommand{\zom}{{\cal Z}(\cH_0, M_n)}
\newcommand{\cP}{{\cal P}}

\newcommand{\G}{\Gamma}
\newcommand{\hX}{\hat{X}}
\newcommand{\Ad}{{Ad}}
\newcommand{\ad}{{ad}}
\newcommand{\loc}{_{\rm loc}}

\def\omi#1{\buildrel { \buildrel{#1}\over{\vee} } \over .}

\def\gone{{\mathbb 1}}
\def\gR{{\mathbb R}}
\def\gQ{{\mathbb Q}}
\def\gK{{\mathbb K}}
\def\gH{{\mathbb H}}
\def\gF{{\mathbb F}}
\def\gC{{\mathbb C}}
\def\gZ{{\mathbb Z}}
\def\gN{{\mathbb N}}
\def\gS{{\mathbb S}}

\def\cA{{\cal A}}
\def\cB{{\cal B}}
\def\cC{{\cal C}}
\def\cD{{\cal D}}
\def\cE{{\cal E}}
\def\cF{{\cal F}}
\def\cG{{\cal G}}
\def\cH{{\cal H}}
\def\cI{{\cal I}}
\def\cJ{{\cal J}}
\def\cK{{\cal K}}
\def\cL{{\cal L}}
\def\cM{{\cal M}}
\def\cN{{\cal N}}
\def\cO{{\cal O}}
\def\cP{{\cal P}}
\def\cQ{{\cal Q}}
\def\cR{{\cal R}}
\def\cS{{\cal S}}
\def\cT{{\cal T}}
\def\cU{{\cal U}}
\def\cV{{\cal V}}
\def\cW{{\cal W}}
\def\cX{{\cal X}}
\def\cY{{\cal Y}}
\def\cZ{{\cal Z}}

\def\tg{{\mathfrak g}}
\def\tm{{\mathfrak m}}
\def\th{{\mathfrak h}}
\def\tA{{\mathfrak A}}
\def\tH{{\mathfrak H}}
\def\tG{{\mathfrak G}}
\def\tX{{\mathfrak X}}
\def\tY{{\mathfrak Y}}
\def\tU{{\mathfrak U}}
\def\tsl{{\mathfrak{sl}}}
\def\tsu{{\mathfrak{su}}}

\def\fA{{\mathfrak A}}
\def\fB{{\mathfrak B}}

\def\Ug{{\cU(\tg)}}
\def\Wg{{\cW(\tg)}}

\def\mrm#1{\mathrm{#1}}
\def\findem{~\hfill$\square$}
\def\finrem{}
\def\finex{~\hfill$\diamondsuit$}
\def\demo{\noindent {\sl Proof: }}
\def\lie{{\mathrm{Lie}}}
\def\hom{{\mathrm{Hom}}}
\def\ker{{\mathrm{Ker\:}}}
\def\der{{\mathrm{Der}}}
\def\aut{{\mathrm{Aut}}}
\def\int{{\mathrm{Int}}}
\def\inv{{\mathrm{inv}}}
\def\bas{{\mathrm{basic}}}
\def\out{{\mathrm{Out}}}
\def\Int{{\mathrm{Int}}}
\def\Bas{{\mathrm{Bas}}}
\def\End{{\mathrm{End}}}
\def\Diff{{\mathrm{Diff}}}
\def\Id{{\mathrm{Id}}}
\def\sign{{\mathrm{sign}}}
\def\diag{{\mathrm{diag}}}
\def\im{{\mathrm{Im\:}}}
\def\tr{{\mathrm{Tr}}}
\def\gr{{\mathrm{gr}}}
\def\op{{\mathrm{op}}}
\def\bic{{\mathrm{Bic}}}
\def\invg{{\mathrm{L-inv}}}
\def\invd{{\mathrm{R-inv}}}
\def\cl{{\mathrm{C}\ell }}
\def\mod{{\mathrm{\ mod\ }}}
\def\fon{{\mathrm{Fon}}}
\def\sp{{\mathrm{Sp}}}
\def\ch{{\mathrm{Ch}}}
\def\ind{{\mathrm{Ind}}}
\def\ensvide{{\emptyset}}
\def\exter{{\textstyle\bigwedge}}
\def\syme{{\textstyle\bigvee}}
\def\exter{{\textstyle\bigwedge}}

\def\vect{{\mathrm{Vect}}}
\def\Re{{\mathrm{Re}}}
\def\Im{{\mathrm{Im}}}

\title{Invariant noncommutative connections}
\author{Thierry Masson$^{\, *}$  \and  Emmanuel Seri\'e$^{\, *, **}$ } 
\maketitle
\begin{center}
  $^{*}$
  Laboratoire de Physique Th\'eorique (UMR 8627)\\
  Universit\'e Paris XI,\\
  B\^atiment 210, 91405 Orsay Cedex, France\\
\end{center}
\begin{center}
  $^{**}$
  Laboratoire de Physique Th\'eorique des Liquides,\\
  Universit\'e Pierre-et-Marie-Curie - CNRS UMR 7600 \\
  Tour 22, 4-\`eme \'etage, Bo\^{i}te 142, \\
  4, Place Jussieu, 75005 Paris, France\\  
\end{center}
\begin{abstract}
In this paper we classify invariant noncommutative connections in the framework of the algebra of endomorphisms of a complex vector bundle. It has been proven previously that this noncommutative algebra generalizes in a natural way the ordinary geometry of connections. We use explicitely some geometric constructions usually introduced to classify ordinary invariant connections, and we expand them using algebraic objects coming from the noncommutative setting. The main result is that the classification can be performed using a ``reduced'' algebra, an associated differential calculus and a module over this algebra. 
\end{abstract}
\vfill
LPT-Orsay 04-54
\newpage
\section*{Introduction}
Symmetries in gauge fields theories play an important role in physics. For instance, they have been used as possible procedures to introduce scalar Higgs fields using dimensional reductions. Solitons, as well as instantons and monopoles, has often been introduced and/or recognized as symmetric solutions of gauge fields equations. Extensive mathematical and physical studies of such symmetries have been proposed, using various approaches, and many examples have been given (see for instance \cite{wang:58,kobayashi:63} for a mathematical point of view, and \cite{forgacs:80,harnad:80,jadczyk:84,hudson:84,hussin:94,brodbeck:96} for a more physical point of view).

On the other hand, gauge fields have been generalized in the noncommutative framework, in some very natural ways. For a large class of examples developed so far, noncommutative connections incorporate at the same time, and without too much arbitrariness, not only ordinary non abelian gauge fields, but also some scalar fields which can be naturally interpreted as Higgs fields (see \cite{dubois-violette:japan99,madore:99} for reviews, and references therein).

In the present paper, we study invariant noncommutative connections in a noncommutative framework which is very strongly connected to ordinary differential geometry.  We take as our starting point a noncommutative algebra equipped with the derivation based differential calculus introduced in \cite{dubois-violette:88}. This algebra is the algebra of endomorphisms of a $SU(n)$-vector bundle. Its structure and its relations to ordinary differential geometry have been extensively studied in \cite{dubois-violette:98,masson:99}, where it has been proven to play a very similar role to a $SU(n)$-principal fiber bundle. In this framework, noncommutative connections extend in a very natural and tractable way ordinary connections on the underlying principal fiber bundle. This permits us to generalize some of the analysis performed in previous works about invariant (ordinary) connections, in particular it is possible to use explicitely some of the geometric constructions related to their classification. Moreover, mixing those geometrical tools and the algebraic approach underlying noncommutative geometry, the classification of invariant noncommutative connections forces us to introduce mathematical objects which are more natural than in the ordinary case.

We have tried to make this paper as self-contained as possible. It is organized as follow. In a first section, we summarize previous works about classification of invariant (ordinary) connections. These results are not new, but we have tried to make a synthesis of the main approaches to this problem. In particular, we introduce there some geometrical constructions which are extensively used later. A second section is devoted to the noncommutative framework. There, we expose the main results about the algebra we will consider, and we try to explain its relations to ordinary differential geometry. New results about the relations between this noncommutative geometry and the underlying ordinary geometry are exposed. Then comes the section which is the main part of this paper. We expose the classification of invariant noncommutative connections, emphasing what is common to the ordinary case and what is new.  In particular, we show that the classification can be performed using objects inspired by noncommutative geometry: a ``reduced'' algebra, an associated differential calculus and a module over this algebra. At last, we study two important examples. One of our motivations for this work was to classify the degrees of freedom of spherically symmetric fields involved in noncommutative models. In the first example, we implement this spherical symmetry in a noncommutative situation which would correspond to an ordinary (trivial) $SU(2)$-principal fiber bundle. The results obtained lead to a natural generalization of the so called Witten's anzatz~\cite{witten:77,forgacs:80}. The second example is a purely noncommutative situation based on a matrix algebra.

\section{Invariant connections on principal fiber bundles}\label{construction_1}
In this section, we would like to summarize previous works on invariant connections in ordinary differential geometry. Firstly, we introduce some notations and some general geometrical constructions which are useful to characterize invariant connections. Two approaches are then proposed: a global one, investigated in~\cite{hudson:84,jadczyk:84,harnad:80}, and a ``local'' one, investigated in~\cite{forgacs:80,brodbeck:96}. The constructions presented here will be used again in section~\ref{invar-nonc-conn} where we characterize noncommutative invarant connections.

\subsection{Reduction of fiber bundles}\label{reduction}
Let us introduce the notations and the hypothesis we make. The analysis presented here is essentially based on the work of Jadczyck {\it et al}~\cite{jadczyk:84}.

We consider a principal fiber bundle $\fiber{H}{E}{M}$, denoted by the following diagram of fibrations
\begin{align*}
  \lfiber[\pi]{H}{E}{M} \ ,
\end{align*}
with structure group $H$. We then consider a compact Lie group $G$ which acts on the left on $E$. We denote this action by $G \circlearrowright E $.  We naturally assume that the two actions $G \circlearrowright E $ and $E \circlearrowleft H$ do commute.  In other words, $G$ will be considered in the following as a subgroup of $\aut(E)$, the group of automorphisms of $E$. The fiber bundle $E$ is called a $G$-symmetric fiber bundle. Then the projection $\pi$ induces an action $G \circlearrowright M$ which is characterized by the following diagram
\begin{align*}
  \xymatrix{0 \ar[r]& \int(E) \ar[r] & \aut(E) \ar[r] & \out(E) \simeq \aut(M)
    \ar[r] & 0  \\
    &&&G \ar@{_(->}[ul]\ar[u]&} 
\end{align*}
A natural problem to consider at this stage is to try to classify all the possible lifts of an action $G \circlearrowright M$ to an action $G \circlearrowright E$. This problem is for instance investigated in~\cite{hussin:94}. We will not touch upon it in the present paper.

We further require the action of $G$ to be simple (see~\cite{harnad:80,jadczyk:84}), which means that we assume $M$ has the following  fiber bundle structure 
\begin{align*}
  \lfiber{G/G_0}{M}{M/G} \ . 
\end{align*}
Then, by hypothesis, the quotient space $M/G$ is a smooth manifold and the fibers are all isomorphic to the homogeneous space $G/G_0$, where $G_0$ is a subgroup of $G$. All isotropy groups for the action $G \circlearrowright M$ are isomorphic to $G_0$, which will denote, once for all, such a chosen reference isotropy group.

Consider now the space $P = \{ x \in M,\ G_x=G_0 \}$, where $G_x$ is the isotropy group associated to any point $x \in M$. It can be shown that $P$ is a principal fiber bundle with structure group $N(G_0)/G_0$, where $N(G_0)$ is the normalizer of $G_0$ in $G$. This fiber bundle is denoted by 
\begin{align*}
  \lfiber{N(G_0)/G_0}{P}{M/G} \ .
\end{align*}
One can then consider the fiber bundle \fiber{G/G_0}{M}{M/G} as an associated fiber bundle to the principal fiber bundle \fiber{N(G_0)/G_0}{P}{M/G} for the natural left action $N(G_0)/G_0 \circlearrowright G/G_0$.
 
A similar construction can be performed on the space $E$ on which the group $S = G \times H $ acts on the right by
  \begin{align*}
    G\times H\times E  & \longrightarrow  E \\
      (g,h,p) & \longmapsto g^{-1} p h 
  \end{align*}
First note that at each point $p\in E$ there exists a canonical homomorphism
\begin{align*}
  \lambda_p  :  G_{\pi (p)} \longrightarrow   H
\end{align*}
defined by the relation $g_0 \cdot p = p \cdot \lambda_p (g_0)$ for all $g_0 $ in $G_{\pi (p)} $. The isotropy group $S_p$ of a point $p$ in $E$ for the action $E \circlearrowleft S$ can be completely characterized, and a straightforward computation shows that $S_p = \{(g_0,\lambda_p(g_0)) / \ g_0 \in G_{\pi (p)} \} $. Then $S_p$ is isomorphic to a generic group $S_0$ for any $p$ in $E$, where $S_0$ is the isotropy group of a certain point $p_0$ in $E$. So $E$ inherits the following fiber bundle structure
\begin{align*}
  \lfiber{S/S_0}{E}{M/G} \ .
\end{align*}
This means that the action of $S$ on $E$ is also simple. Using the same approach as before, one can see that $E$ contains the principal fiber sub-bundle  $Q= \{ p \in E,\ S_p=S_0 \}$ given by the diagram of fibrations
\begin{align*}
\lfiber{N(S_0)/S_0}{Q}{M/G} \ ,
\end{align*}
where $N(S_0)$ is the normalizer of $S_0$ in $S$. Notice that on $Q$ the application $\lambda_p$ is independent of the point $p \in Q$, and  we denote it by  $\lambda: G_0 \to H$.

The restriction of the projection $\pi$ to $Q$ will be called $ \pi_{Q}$. It is obvious that $\pi (Q)  \subset P$. The kernel of $\pi_Q$ is isomorphic to $Z_0=Z(\lambda(G_0),H)$,  the centralizer of $\lambda(G_0)$ in $H$, and 
\begin{align*}
\lfiber[\pi_Q]{Z_0}{Q}{\pi(Q)}
\end{align*}
is a principal fiber bundle. By the very definition one has
\begin{align*}
  N(S_0)= \{(g,h) \in S / \  g \in N(G_0), \
  h^{-1} \lambda(g_0) h = \lambda(g^{-1} g_0 g), \forall g_0 \in G_0 \} \ .
\end{align*}
There is then a natural inclusion  of $Z_0$ in $ N(S_0)/S_0$ given by the following composition of maps
\begin{align*}
  \xymatrix@1@R=7pt@M=6pt{ {Z_0 \simeq \{ e\} \times Z_0 } \ar@{^{(}->}[r] & {N(S_0)} \ar[r] & {N(S_0)/S_0} } \ .
\end{align*}
Furthermore  $Z_0$ is a normal subgroup in $ N(S_0)/S_0$, and one can finally show that $  (N(S_0)/S_0)/Z_0$ is a subgroup of $ N(G_0)/G_0$.

All the previous constructions can be summarized in the following commutative diagram
{
\footnotesize
\begin{equation}
\raisebox{.5\depth-.5\height}{%
\xymatrix@R=7pt@C=3pt@M=6pt{%
Z_0 \ar@{^{(}->}[rrr] \ar@{^{(}->}[ddrr] \ar@{=}[ddd] & & & N(S_0)/S_0 \ar@{>>}[rrr] \ar@{^{(}->}[ddrr] \ar'[dd][ddd] & & & (N(S_0)/S_0) / Z_0 \ar@{^{(}->}[dr] \ar'[dd][ddd] & & \\
& & & & & & & N(G_0)/G_0 \ar@{^{(}->}[dr] \ar'[d][ddd] & \\
& & H \ar@{^{(}->}[rrr] \ar@{=}[ddd] & & & S/S_0 \ar@{>>}[rrr] \ar[ddd] & & & G/G_0 \ar[ddd] \\
Z_0 \ar'[rr][rrr] \ar@{^{(}->}[ddrr] & & & Q \ar@{>>}'[rr][rrr]^-{\pi_Q} \ar@{^{(}->}[ddrr] \ar@{>>}'[dd][ddd] & & & \pi(Q) \ar@{^{(}->}[dr] \ar@{>>}'[dd][ddd] & & \\
& & & & & & & P \ar@{^{(}->}[dr] \ar@{>>}'[d][ddd] & \\
& & H \ar[rrr] & & & E \ar@{>>}[rrr]^-{\pi} \ar@{>>}[ddd] & & & M \ar@{>>}[ddd] \\
& & & M/G \ar@{=}'[rr][rrr] \ar@{=}[ddrr] & & & M/G \ar@{=}[dr] & & \\
& & & & & & & M/G \ar@{=}[dr] & \\
& & & & & M/G \ar@{=}[rrr] & & & M/G \\%
}}\label{diagram}
\end{equation}
}
In this diagram, some arrows represent true applications and other arrows are part of diagrams of fibrations, most of them explicitly given before. Some horizontal arrows correspond to the action of $H$ (or subgroups of $H$) and some vertical arrows correspond to actions of groups related to $G$ and $S$. One can verify that the kernel of the projection $\pi(Q) \to M/G $ is isomorphic to $ (N(S_0)/S_0)/Z_0$.

\subsection{Invariant connections}\label{inv_con}

The action  $G \circlearrowright E$ induces an action of $G$ on the space $\Omega^1(E)$ of 1-forms over $E$. Because the actions of $G$ and $H$  commute, this action extends naturally to an action on the affine space of connections on $E$ included in the space $\Omega^1(E) \otimes \cH$, where $\cH$ is the Lie algebra of $H$. For any $\omega \in \Omega^1(E) \otimes \cH$ and any $g \in G$, we denote this action by $\omega^g= g^{*} \omega$. We are now interested to characterize the  $G$-invariant connections, those which satisfy $\omega^g=\omega$ for any $g \in G$ .

In order to do that, it is convenient to make some natural decompositions of the tangent spaces to the various manifolds introduced previously. These decompositions are performed along the different actions that these spaces support. Let us introduce  in the following table the notations for the Lie algebras corresponding to the groups introduced so far
\begin{displaymath}
  \begin{array}{|c|c|c|c|c|c|c|c|c|c|} 
    \hline
    \text{Group} &    G & H & N(G_0) & G_0 & N(G_0)/G_0 & Z_0 & S=G\times H & S_0 & N(S_0) \\
    \hline
    \text{Lie Algebra} &  \cG &  \cH & \cN_{0}  & \cG_0 & \cK & \zo & \cS= \cG \oplus \cH & \so & \cN_{\so} \\
    \hline
  \end{array}
\end{displaymath}
Let,
\begin{align*}
  \cG &= \cN_0 \subsetplus \cL &
  & \text{and} &
  \cH &= \zo \subsetplus \cM \ ,
\end{align*}
be some reductive decompositions of Lie algebras\footnote{A  decomposition of a Lie algebra  $\mathfrak{g}=\mathfrak{h} \subsetplus \mathfrak{l} $ is reductive when  $\mathfrak{h} \subset \mathfrak{g} $ is a sub Lie algebra and $\mathfrak{l}$  a reductive complementary subspace, i.e   $[  \mathfrak{h} , \mathfrak{l} ] \subset \mathfrak{l}$} which we suppose to be also  orthogonal decompositions of vector spaces for the Killing metrics. It is easy to show that there is an orthogonal decomposition of Lie algebras
\begin{align*}
\cN_0= \cG_0 \oplus \cK
\end{align*}
Then, by the very definitions, one has
\begin{align*} \so =  \{ (X_0, \lambda_* X_0) / X_0 \in \cG_0 \} \ , \end{align*}
 where $\lambda_*: \cG_0 \to \cH$ is the tangent application to $\lambda: G_0 \to H$, which implies that $\so$ is isomorphic to $\cG_0$. Using this identification, one can easily show that 
\begin{align*}
  \cN_{\so} =\so \oplus \cK \oplus \zo 
\end{align*}
is an orthogonal decomposition of Lie algebras. 
In fact any element $(X,\xi) \in \cN_{\so}\subset \cG \times \cH$ can be written in the form $(X,\xi)=(X_0+X_{\cK},\lambda_*X_0 +\xi_{\zo})$, where $X_0 \in \cG_0, X_{\cK} \in \cK$ and $\xi_{\zo}\in\zo$.

With these decompositions and the induced maps of the group actions on manifolds at the level of Lie algebras and tangent spaces, we can decompose the different tangent spaces of the bundles introduced previously. We then get an infinitesimal version of diagram~(\ref{diagram})
{\footnotesize
  \begin{align*}
    \raisebox{.5\depth-.5\height}{%
      \xymatrix@R=7pt@C=3pt@M=6pt{       
        {\zo} \ar@{^{(}->}[rrr] \ar@{^{(}->}[ddrr] \ar@{=}[ddd] & & &  {\cK \oplus \zo} \ar@{>>}[rrr] \ar@{^{(}->}[ddrr] \ar@{^{(}->}'[dd][ddd] & & & \cK \ar@{  =}[dr] \ar@{^{(}->}'[dd][ddd] & & \\
        & & & & & & & \cK \ar@{^{(}->}[dr] \ar@{^{(}->}'[d][ddd] & \\
        & & {\zo \subsetplus \cM} \ar@{^{(}->}[rrr] \ar@{=}[ddd] & & & { (\cK \subsetplus \cL) \oplus (\zo \subsetplus \cM)} \ar@{>>}[rrr] \ar@{^{(}->}[ddd] & & & \cK \subsetplus \cL \ar@{^{(}->}[ddd] \\
        {\zo} \ar@{^{(}->}'[rr][rrr] \ar@{^{(}->}[ddrr] & & &{ T_q Q} \ar@{>>}'[rr][rrr]^-{\pi_{Q *}} \ar@{^{(}->}[ddrr] \ar@{>>}'[dd][ddd] & & & {T_x \pi(Q) } \ar@{^{(}->}[dr] \ar@{>>}'[dd][ddd] & & \\
        & & & & & & & {T_x P} \ar@{^{(}->}[dr] \ar@{>>}'[d][ddd] & \\
        & & {\zo \subsetplus \cM} \ar@{^{(}->}[rrr] & & & {T_q E} \ar@{>>}[rrr]^-{\pi_*} \ar@{>>}[ddd] & & & {T_x M} \ar@{>>}[ddd] \\
        & & & {T_{[x]} M/G} \ar@{=}'[rr][rrr] \ar@{=}[ddrr] & & & {T_{[x]} M/G} \ar@{=}[dr] & & \\
        & & & & & & & {T_{[x]} M/G} \ar@{=}[dr] & \\
        & & & & & {T_{[x]} M/G} \ar@{=}[rrr] & & &  {T_{[x]} M/G} \\%
        }}%
    \end{align*}
     }
 for a point  $q\in Q \subset E$ with $\pi(q)=x$. Hence we have
 \begin{align}
   T_q E= T_q Q \oplus\cL^{Q}_{|q} \oplus \cM^Q_{|q} \ , \label{T_decomposition}
 \end{align}
  where $ \cL^Q_{|q}$ (resp. $\cM^Q_{|q}$) is the subvector space obtained from the values $X_q^E$ at $q\in E$ of the fundamental vector fields $X^E$ on $E$ associated to vectors $ X\in \cL$ (resp. $X \in \cM$).

  Let us now consider a $G$-invariant connection 1-form $\omega \in \Omega^1(E)\otimes \cH$. We are interested to characterize the degrees of freedom of $\omega$. First, note that it is possible to study $\omega$ only at points in $Q$ owing to the fact that by construction $G \cdot Q = E $. Then for any $q \in Q$, $\omega_{|q}$  can be evaluated on the 3 vector spaces $T_q Q$, $\cL^Q_{|q}$ and $\cM^Q_{|q} $
 \begin{itemize}
 \item The restriction to $\cM^Q_q \subset \cH^Q_q $ is fixed by the relation $\omega_{|q} (X_q^E) = X$, for any $X \in \cM$. So there is no degree of freedom in this direction.
 \item The restriction to $T_q Q$ gives a  1-form $\mu$ defined by  $\mu (X)= \omega (X)$, for any $X \in TQ $. It satisfies the following equivariance property
   \begin{align*}
     &\R^{*}_{(g,h)} \mu = Ad_{h^{-1}} \mu &
     &\forall (g,h) \in N(S_0) 
   \end{align*}
   where $\R_{(g,h)}$ is the right action of $N(S_0)$ on $Q$. Considering this equivariance property for an element $(g_0, \lambda (g_0) ) \in S_0$, one can show that $\mu$ is a $\zo$-valued 1-form. Together with the equivariance property restricted to $Z_0$, this implies in particular that $\mu$ is a connection on the principal fiber bundle $ \fiber{Z_0}{Q}{\pi (Q)}$.
 \item The restriction to $\cL^Q_q $ induces a map
   \begin{align*}
     \psi_q:    \cL   & \to  \cH   \\
     X& \mapsto \psi_q (X) = \omega_q (X_q^E)
   \end{align*}
   It satisfies the following equivariance property:
   \begin{align*}
     &\ Ad_h \circ \psi_q \circ Ad_{g^{-1}} = \psi_{g q h^{-1}} &
     &    \forall (g,h) \in S 
   \end{align*}
   Then, for any $(g,h) \in S_0$, one has $ Ad_{\lambda(g_0)} \circ \psi_{q} \circ Ad_{g_0^{-1}} =\psi_q $. The  equivariant map   $q \mapsto \psi_q$ from $Q$ to $\cF$,  defines a section of the associated vector bundle  $ F^{\cL} = Q \times_{N(S_0)/S_0} \cF_{\cL}$, where the fiber is defined to be the vector space of covariant maps
   \begin{align*}
     \cF_{\cL}= \{\ell : \cL\to \cH  , Ad_{\lambda (g_0)} \circ \ell \circ Ad_{g_0^{-1}} = \ell  \}
   \end{align*}
 \end{itemize}

 So $\omega$ is completely characterized by the two objects $\mu$ and $\psi$ described above. Notice that $\mu$ and $\psi$ can be interpreted as genuine objects on fiber bundles related to the two principal fiber bundle structure on $Q$, either in the ``horizontal'' direction for $\mu$ or in the ``vertical'' direction for $\psi$ on diagram~(\ref{diagram}). It is possible to make reference to only one of these principal fiber bundles structures. In order to do that, one needs a reference connection $A$ on the principal fiber bundle 
 \begin{align*}
 \lfiber{(N(S_0)/S_0)/Z_0}{\pi (Q)}{M/G}
 \end{align*}
Then one can make $\mu$ in one-to-one correspondence with a couple  $(B, \alpha)$ where 
 \begin{itemize}
 \item $B$ is a connection on the principal fiber bundle \fiber{N(S_0)/S_0}{Q}{M/G}.
 \item $\alpha$ is a section of the vector bundle $F^{\cK}$,  where  $F^{\cK} = Q \times_{N(S_0)/S_0} \cF_{\cK} $ is associated to \fiber{N(S_0)/S_0}{Q}{M/G}. The vector space $\cF_{\cK}$ is defined by
 \begin{align*}
 \cF_{\cK}= \{ k : \cK \to \cH  , Ad_{\lambda (g_0)} \circ k \circ  Ad_{g_0^{-1}} =  k  \}
 \end{align*}
 \end{itemize}
 (Notice the similarity between $F^{\cL}$ and $F^{\cK}$.) \\
 The correspondence between  $\mu$ and the pair $(B,\alpha)$ is given explicitely by the relations
 \begin{align*}
   \left\{
     \begin{array}{ll}
       B & =  \mu + \pi^* A  - \mu (\pi^* A)^Q  \\
       \alpha_q &= \mu_q |_{\cK_q} 
     \end{array} \right.
 \end{align*}
 In particular we have that $pr_\cK B = \pi^* A $, where $pr_\cK$ is the orthogonal projection from $\cK \oplus \zo $ to $\cK$. We refer to~\cite{jadczyk:84} for further details and the works by Coquereaux {\it et al}~\cite{coquereaux:85} for the link with Kaluza-Klein theories.

 \subsection{Relation with Wang's approach}\label{wang_approach}
 Because we have assumed that the action of the group $G$ is simple, the space $M$ is locally isomorphic to the product space $ M/G \times G/G_0$. The study of invariant connections is greatly simplified if one considers the space $M$ to be exactly equal to the space  $M/G \times G/G_0$. This is also equivalent to restrict the study only to local objects around an orbit of $G$ in $M$. So, in the following, we will assume that $M=M/G \times G/G_0$, and we will expose the main results obtained in~\cite{harnad:80,forgacs:80,brodbeck:96}. This is what we call the ``local'' approach.

 The simpler structure of the space $M$ allows us to do a construction similar to the one performed  previously, replacing the space $P$ by the space $M/G$. This simplifies  the bundle structure in the  $G$'s directions and also greatly simplifies the decomposition of invariant connections. Furthermore, it is possible to classify the $G$-symmetric fiber bundles, and the results in this special case can be compared more easily with the ones obtained by Wang~\cite{wang:58}.

 First, because of the decomposition $M = M/G \times G/G_0$,  one can imbed $M/G$ into $M$, identifying it with $M/G \times \{e G_0\}$. Then, $G$-symmetric principal $H$-bundles can be classified by pairs $([\lambda],\tQ)$, where $[\lambda]$ is a conjugacy class of homomorphisms $\lambda : G_0 \to H $ for the action of $G$ on $G_0$ by conjugation, and $\tQ$ is a principal fiber bundle over $M/G$ with structure group $Z_0= Z(\lambda(G_0),H)$. 

 Indeed, one can construct a pair $([\lambda],\tQ)$ from a $G$-symmetric principal $H$-bundle $E$ over $M = M/G \times G/G_0$ considering the restriction  $E_{|M/G}$ of $E$ over $M/G$. Then, define $\tQ= \{ p \in E_{|M/G} / \lambda_p=\lambda \}$, for $\lambda$ a chosen reference map $\lambda_{p_0}$ for a $p_0 \in E$.

 Conversely, one can associate a $G$-invariant principal $H$-bundle to any pair $([\lambda],\tQ)$. In order to do that, it is convenient to introduce the following diagram of fibrations
 \begin{align*} 
 \lfiber{Z_0 \times G_0}{Q'=\tQ \times G}{M/G \times G/G_0} 
 \end{align*}
 which defines a principal fiber bundle $Q'$ for the action $(z,g_0,\tq,g)\mapsto(\tq \cdot z_0, g \cdot g_0)$. Consider now the following left action of $Z_0 \times G_0$ on $H$ defined by\footnote{notice that the induced actions of the subgroups $Z_0$ and $G_0$ commute.}
 \begin{align*}
   \rho:  Z_0 \times G_0 \times H  &\longrightarrow H \\
   (z,g_0,h) &\longmapsto z \cdot \lambda(g_0) \cdot h 
 \end{align*}
 Denote by $\tE= Q' \times_{(Z_0 \times G_0)} H$ the associated fiber bundle to $Q'$ with fiber $H$ for this action. It can be shown that $\tE$ is a $G$-invariant principal $H$-bundle characterized by the following commutative diagram\footnote{Some arrows are part of diagram of fibrations.}
 \begin{align}
\raisebox{.5\depth-.5\height}{%
  \xymatrix{
    {}& {Z_0 \times G_0} \ar[d] &  {Z_0 \times G_0} \ar[d] \\
    {H} \ar[r] & {Q' \times H} \ar[r]^-{pr_1} \ar@{->>}[d]_-{\Psi} & {Q'=\tQ \times G} \ar@{->>}[d] \\
    {H} \ar[r]  & {\tE} \ar[r]^-{\tilde{\pi}} & {M=M/G \times G/G_0}}}
\label{fibre_associe}
\end{align}
 where $Q'\times H$ is also a $G$-invariant principal $H$-bundle for the horizontal structure.  

 It is easy to prove that the composition of these two maps, $E\mapsto ([\lambda],\tQ)$ and $ ([\lambda],\tQ)\mapsto \tE$, gives us a map $E \mapsto \tE$, for which $E$ and $\tE$ are isomorphic $G$-invariant principal $H$-bundles. Indeed, an isomorphism between $E$ and $\tE$ is given explicitely by the following relation: to any point $\Psi(\tq,g,h) \in \tE$, associate the point $g \cdot \tq \cdot h \in E$ where $\tq\in \tQ$ is considered to be in $E$.

 Using this isomorphism, it is possible to map a $G$-invariant connection on $E$ to a $G$-invariant connection $\omega $ on $\tE$. Now, owing to the fact that the projection $\Psi$ of the principal $(Z_0 \times G_0)$-bundle $Q'\times H$ is also a $G$-equivariant map of $G$-symmetric principal $H$-bundles, one can show (see~\cite{harnad:80,brodbeck:96}) that $\pso$ can be written in the generic form
 \begin{align}
  \pso_{|(\tq,g,h)} = Ad_{h^{-1}} ( \Lambda_{|\tq} \circ \theta^G_{~|g} + \tom_{|\tq}) +\theta^H_{~|h} \ ,  
  \label{class_decomposition}
\end{align}
where $\tom$ is a connection 1-form on $\fiber{Z_0}{\tQ}{M/G}$, $\theta^G$ and $\theta^H$  are the usual Cartan 1-form on $G$ and $H$ respectively, and $ \Lambda \in C^{\infty}(\tQ)\otimes \cG^* \otimes \cH $  satisfies the equivariance property:
\begin{align*}
  &\R_{z_0}^* \Lambda = Ad_{z_0} \Lambda &
  & \forall z_0 \in Z_0
\intertext{and the two relations}
  &Ad_{\lambda(g_0)} \circ  \Lambda  \circ Ad_{g_0^{-1}} =  \Lambda  &
  &\forall g_0 \in G_0 \\
  &  \Lambda_{\tq}(X_0) = \lambda_*(X_{0}) &
  &\forall X_0 \in \cG_0 \ \text{and} \ \forall \tq \in \tQ 
\end{align*}
Using standard technics in differential geometry, this equivariant map $\Lambda: \tQ \to \cG^*\otimes\cH$  characterizes a section of a vector bundle over $M/G$, with fiber $\cG^* \otimes \cH$, associated to $\tQ$ for the adjoint action of $Z_0$ on $\cH$.

\bigskip

We would like to conclude this section by the following remark. In the two situations presented here, the ``global'' one and the ``local'' one, it is possible to characterize the $G$-invariant connections on the $G$-symmetric principal $H$-bundle $E$ using geometric objects related to the quotient space $M/G$, and not to the whole space $M$. Nevertheless, for the ``global'' approach, this requires an extra arbitrary connection $A$.

\section{The noncommutative differential calculus}\label{ncdc}

In this section, we introduce the derivations based differential calculus defined for any associative algebra~\cite{dubois-violette:88} and describe more precisely this calculus for the algebra of endomorphisms of a complex vector bundle introduced and studied in~\cite{dubois-violette:98,masson:99}. Some new results extending these previous studies are presented here. It is in the framework of this noncommutative geometry that we will study $G$-invariant noncommutative connections in the next section. The notion of $G$-invariant noncommutative connection is introduced at the end of this section, as well as some supplemental mathematical structures that will be used later.

\subsection{Derivation-based differential calculus}

In the following, $\fA$ will denote an associative algebra with unit. Then the vector space $\der(\fA)$ of derivations of $\fA$ is a Lie algebra and a module over the center $\cZ(\fA)$ of $\fA$. The vector space of inner derivations, $\Int(\fA)$, is a Lie ideal and a $\cZ(\fA)$-submodule. The quotient $\der(\fA)/\Int(\fA)$ will be denoted by $\out(\fA)$. This is a Lie algebra and a module over $\cZ(\fA)$. 

Define the complex $\underline{\Omega}_\der(\fA)$ to be the set of 
$\cZ(\fA)$-multilinear antisymmetric maps from $\der(\fA)$ to $\fA$. It is 
naturally a $\gN$-graded algebra on which one can define a differential $\hd$ 
(of degree $1$) by setting, for any derivations $X_1, \dots, X_{n+1}$ and any $\omega \in \underline{\Omega}^n_\der(\fA)$
\begin{eqnarray}
\hd\omega(X_1, \dots , X_{n+1}) &=& \sum_{i=1}^{n+1} (-1)^{i+1} X_i
\omega( X_1, \dots \omi{i} \dots, X_{n+1}) \nopagebreak\nonumber\\
\nopagebreak & & + \sum_{1\leq
i < j \leq n+1} (-1)^{i+j} \omega( [X_i, X_j], \dots \omi{i} \dots
\omi{j} \dots , X_{n+1})
\label{differential}
\end{eqnarray}
In the following, for all the associative algebras we will consider, this graded differential algebra $(\underline{\Omega}_\der(\fA), \d)$ coincides with the smallest differential subalgebra of $\underline{\Omega}_\der(\fA)$ generated by $\fA$, which is usually denoted by 
$\Omega_\der(\fA)$. 

Let $\cG$ be a Lie subalgebra of $\der(\fA)$. Then $\cG$ defines a natural operation in the sense of H.~Cartan~\cite{cartan:51} on $(\underline{\Omega}_\der(\fA), \d)$. Indeed, for any $X\in \cG$ and $n\geq 1$, let us introduce
\begin{equation*}
i_X : \underline{\Omega}^n_\der(\fA) \rightarrow \underline{\Omega}^{n-1}_\der(\fA)
\end{equation*}
by
\begin{equation*}
(i_X \omega)( X_1, \dots , X_{n-1}) = \omega (X, X_1, \dots , X_{n-1})
\end{equation*}
for any $\omega \in \underline{\Omega}^n_\der(\fA)$ and $X_i \in \der(\fA)$. This interior product is defined to be $0$ on $\underline{\Omega}^0_\der(\fA)=\fA$. It is easy to show that $i_X$ is a graded derivation of degree $-1$ on $\underline{\Omega}_\der(\fA)$. The application
\begin{equation*}
L_X = i_X \hd + \hd i_X  : \underline{\Omega}^n_\der(\fA) \rightarrow \underline{\Omega}^{n}_\der(\fA)
\end{equation*}
defined for any $n\geq 0$ is then a graded derivation of degree $0$ on the graded algebra $\underline{\Omega}_\der(\fA)$. This is the Lie derivative  associated to the operation of $\cG$ on $(\underline{\Omega}_\der(\fA), \hd)$. One can compute the usual relations
\begin{align*}
i_X i_Y + i_Y i_X &= 0 & L_X i_Y - i_Y L_X &= i_{[X,Y]} \\
L_X L_Y - L_Y L_X &= L_{[X,Y]} & L_X \hd - \hd L_X &= 0
\end{align*}
In the same way, it is possible to define a natural Cartan operation of $\cG$ on $(\Omega_\der(\fA), \hd)$.

With such an operation of $\cG$ on $(\underline{\Omega}_\der(\fA), \hd)$, one can introduce the basic subspace of $\underline{\Omega}_\der(\fA)$, which is the common kernel of all the $i_X$ and $L_X$ for all $X \in \cG$. This basic subspace can be shown to be a graded differential subalgebra. The common kernel of all the $L_X$ for all $X \in \cG$ is called the invariant subspace of the operation. This is also a graded differential subalgebra.

In the case where $\fA$ is the algebra $C^\infty(M)$ of smooth complex-valued functions on a finite dimensional regular manifold $M$, $(\Omega_\der(\fA), \hd)$ is just the de~Rham complex $(\Omega(M), \d)$ and $\der(C^\infty(M))=\Gamma(TM)$ is the ordinary Lie algebra of vector fields on $M$.

Let us consider the case where $\fA$ is the algebra $M_n:=M_n(\gC)$ of $n\times n$ complex matrices~\cite{dubois-violette:90}. This algebra has only inner derivations, and the Lie algebra $\der(M_n) = \Int(M_n)$ can be identified with the Lie algebra $\tsl_n:=\tsl(n,\gC)$. One can show that
\begin{align}
  \Omega_\der(M_n) &\simeq M_n \otimes \exter \tsl_n^\ast
\label{soudure}
\end{align}
where $\tsl_n^\ast$ is the dual of $\tsl_n$. We denote by $\d'$ the differential on this complex.

In this situation, there exists a particular $1$-form $\theta$ defined by
\begin{align*}
  i\theta : \der(M_n) & \longrightarrow \tsl_n\\
  ad_{\gamma} & \longmapsto \gamma - \frac{1}{n} \tr (\gamma)\gone
\end{align*}
for any $\gamma \in M_n$. This $1$-form satisfies to the relation
\begin{align*}
 \d' i\theta - (i\theta)^2 = 0
\end{align*}
and for any $\gamma \in M_n = \Omega^0_\der(M_n)$, one has $\d' \gamma = [i\theta, \gamma]$. This 1-form $\theta$ can also be viewed as a kind of fundamental 1-form in this noncommutative space, which permits one to explicitely identify the Lie algebras $\der(M_n)$ and $\tsl_n$.

Let us now mix the two previous examples in a trivial way, taking the matrix valued functions on a manifold $M$:  $\fA = C^\infty(M)\otimes M_n$. The derivations based differential calculus for this algebra has been studied in~\cite{dubois-violette:90:II}. Here are the main results. The center of the algebra $\fA$ is exactly $C^\infty(M)$, and the Lie algebra of derivations $\der(\fA)$ splits canonically as a $C^\infty(M)$-module into
\begin{equation}
\label{decder1}
\der(\fA) = [\der(C^\infty(M))\otimes \gone ] \oplus [ C^\infty(M) \otimes \der(M_n) ] 
\end{equation}
This implies the canonical decomposition for the complex of forms
\begin{align*}
  \Omega_\der(\fA) = \Omega(M) \otimes \Omega_\der(M_n)
\end{align*}
The differential $\hd$ on $\Omega_\der(\fA)$ is the sum $\hd = \d + \d'$ where $\d$ and $\d'$ has been defined in the two previous examples. The $1$-form $\theta$ is well defined in $\Omega^1_\der(\fA)$ if we extend it on $\der(\fA)$ by zero on the $\Gamma(TM)$ terms. 

\subsection{Algebra of endomorphisms of a vector bundle}\label{algebra-endom-vect}

Let us now consider a non trivial version of the previous example. Let $\cE$ be a $SU(n)$-vector bundle over a regular finite dimensional smooth (i.e. paracompact, etc...) manifold $M$ equipped with an hermitian structure. We denote by $\End(\cE)$ the fiber bundle of endomorphisms of $\cE$. The sections of this fiber bundle in matrix algebras define a unital algebra, which we denote by $\fA$. The hermitian structure gives a natural involution on this algebra, denoted by $S \mapsto S^\ast$. The center of this algebra is exactly $C^\infty(M)$, identifying $f\in C^\infty(M)$ with $f\gone \in \fA$. The trace map and the determinant, defined on each fiber of $\End(\cE)$, give natural maps
\begin{align*}
  \tr : \fA \rightarrow C^\infty(M) \mbox{ and } \det : \fA \rightarrow  C^\infty(M)
\end{align*}
By restriction to the center,  there is also a natural map 
\begin{align}
  \rho : \der(\fA) \rightarrow \der(C^\infty(M)) = \Gamma(TM)
\label{rho}
\end{align}
This map is the quotient map in the short exact sequence of Lie algebras and $C^\infty(M)$-modules
\begin{equation}
\label{sesder}
\xymatrix@1{ {0} \ar[r] & {\Int(\fA)} \ar[r] & {\der(\fA)} \ar[r]^-{\rho} & {\out(\fA) \simeq  \Gamma(TM)} \ar[r] & {0}}
\end{equation}
This short exact sequence generalizes the decomposition (\ref{decder1}) in the trivial case. Notice that in the non trivial case, one cannot split canonically this short exact sequence of $C^\infty(M)$-modules. 

For any derivation $\cX \in \der(\fA)$, let us denote by $X\in \Gamma(TM)$ the associated vector field on $M$. The $1$-form $i\theta$ defined in the two previous examples is well defined here on $\Int(\fA)$ only, by the relation
\begin{align*}
  i\theta(ad_\gamma) = \gamma - {1\over n}\tr(\gamma) \gone
\end{align*}
for any $\gamma\in \fA$. In the following, for any inner derivation $ad_\gamma$, we suppose that the element $\gamma$ is traceless. It can be considered as a section of the fiber bundle of traceless endomorphisms of $\cE$. We denote by $\fA_0$ the space of traceless elements in $\fA$. The Lie subalgebra $\Int(\fA)$ operates in the sense of H. Cartan on the differential complex $\Omega_\der(\fA)$~\cite{dubois-violette:88}. The horizontal forms for this operation are exactly the differential forms on $M$ with values in $\End(\cE)$, and the basic forms are ordinary differential forms on $M$. In the following, horizontality will refer to this operation.

It was shown in~\cite{dubois-violette:98} that the two differential calculi $\Omega_\der(\fA)$ and $\underline{\Omega}_\der(\fA)$ coincide. We will denote by $\hd$ the differential on $\Omega_\der(\fA) = \underline{\Omega}_\der(\fA)$.

Now, let $\nabla^\cE$ be any connection on $\cE$. Then it was shown in~\cite{dubois-violette:98} that there exists a noncommutative $1$-form $\alpha$ in $\Omega^1_\der(\fA)$ such that any derivation $\cX \in \der(\fA)$ can be decomposed as
\begin{equation}
\label{decder} 
\cX = \nabla_X - ad_{\alpha(\cX)} 
\end{equation}
where $\nabla$ is the naturally associated connection to $\nabla^\cE$ on the fiber bundle  $\End(\cE)$. Indeed, one can define $\alpha$ by the relation $\alpha(\cX)= -i\theta(\cX - \nabla_X)$. We recall that $\nabla$ is the tensor product of the connections  $\nabla^\cE$ on $\cE$ and $\nabla^{\cE^\ast}$ on the dual vector bundle $\cE^\ast$ of $\cE$ where $\nabla^{\cE^\ast}$ satisfies $X \langle \epsilon , e \rangle = \langle \nabla^{\cE^\ast}_X \epsilon , e \rangle + \langle \epsilon , \nabla^\cE_X e \rangle$ for any sections $\epsilon$ of $\cE^\ast$ and $e$ of $\cE$.

The noncommutative $1$-form $\alpha$ takes its values in the traceless elements of $\fA$ and can be considered as an extension of $-i\theta$ to all derivations. One has obviously $\alpha(ad_\gamma) = -\gamma$, with the convention that $\tr(\gamma) = 0$.

This result gives us a splitting of the short exact sequence~(\ref{sesder}) as $C^\infty(M)$-modules. This splitting is not canonical and is only defined through a choice of a connection on $\cE$, by the $C^\infty(M)$-linear map $X\mapsto \nabla_X$ from $\Gamma(TM)$ into $\der(\fA)$. This has to be compared with the usual (commutative) situation where one can interpret a connection as a map from vector fields on $M$ into vector fields on a principal bundle over $M$. These maps will be used and generalized in  \ref{sec:omega_der}.

The algebra $\fA$ plays a similar role to a principal bundle, and the above canonical map $\nabla^\cE \mapsto \alpha$ is an isomorphism of affine spaces from the affine space of $SU(n)$-connections on $\cE$ onto the affine space of traceless antihermitian noncommutative $1$-forms on $\fA$ satisfying $\alpha(ad_\gamma) = -\gamma$. If $R^\cE$ denotes the curvature of $\nabla^\cE$, then one can show that 
\begin{align*}
  R^\cE(X,Y) = \hd\alpha(\cX, \cY) + [\alpha(\cX), \alpha(\cY) ]
\end{align*}
for any $\cX, \cY \in \der(\fA)$, $X,Y$ being their images in $\Gamma(TM)$.  In particular, the expression $\hd\alpha + \alpha^2$ is a horizontal element of $\Omega^2_\der(\fA)$.

Now, the Lie algebra of real derivations on $\fA$ acts naturally on the space of $SU(n)$-connections through the Lie derivative defined on $\Omega_\der(\fA)$. If one restricts this action to inner real derivations, the Lie derivative corresponds to infinitesimal gauge transformations on connections. Indeed, one has
\begin{align*}
  \cL_{ad_\xi} \alpha = -\hd \xi - [\alpha, \xi]
\end{align*}
for any $\xi \in \cA$, with $\tr \xi =0$ and $\xi^\ast + \xi = 0$ ($ad_\xi$ is then a real inner derivation). Such $\xi$'s are exactly the elements of the Lie algebra of the group of gauge transformations on $\cE$.

\subsection{Noncommutative connections}

In the following, we will only consider noncommutative connections for the algebra $\fA$ defined above on the right module $\fA$ itself (This definition could be given for any associative algebra $\fA$). A noncommutative connection is an application 
\begin{align*}
  \widehat{\nabla}_\cX : \fA \rightarrow \fA
\end{align*}
such that $\widehat{\nabla}_\cX(S S') = S \cX(S') + \widehat{\nabla}_\cX(S) S'$ and $\widehat{\nabla}_{f\cX} S = f\widehat{\nabla}_\cX S$ for any $\cX \in \der(\fA), S,S' \in \fA$ and $f\in C^\infty(M)$. The curvature of a noncommutative connection is defined by $\hat{R}(\cX, \cY) S = [ \widehat{\nabla}_\cX, \widehat{\nabla}_\cY ] S - \widehat{\nabla}_{[\cX, \cY]}S$ for any $S \in \fA$ and $\cX, \cY \in \der(\fA)$, which is a right $\fA$-module homomorphism.

Any noncommutative connection $\widehat{\nabla}$ on $\fA$ is completely given by $\widehat{\nabla}_\cX \gone = \omega(\cX)$, where $\omega$ is a noncommutative $1$-form in $\Omega^1_\der(\fA)$. Indeed, one then has
\begin{align*}
  \widehat{\nabla}_\cX S = \cX S + \omega(\cX) S
\end{align*}
for any $S\in \fA$. The curvature of $\widehat{\nabla}$ is then the left multiplication by the noncommutative $2$-form 
\begin{align*}
  \hd\omega (\cX, \cY) + [ \omega(\cX), \omega(\cY) ]
\end{align*}

There is a natural hermitian structure on the right module $\fA$ given by $\langle S, S' \rangle = S^\ast S' \in \fA$. A connection is said to be compatible with an hermitian structure if 
\begin{align*}
  \cX \langle S, S' \rangle = \langle \widehat{\nabla}_\cX S, S' \rangle + \langle S, \widehat{\nabla}_\cX S' \rangle
\end{align*}
for any $S,S' \in \fA$ and real\footnote{A derivation $\cX\in \der(\fA)$ is real if $(\cX a)^\ast=\cX a^\ast$ for any $a \in \fA$} $\cX \in \der(\tA)$.  This compatibility condition is equivalent to  
\begin{align*}
  \omega(\cX)^\ast + \omega(\cX) = 0
\end{align*}
for any real $\cX \in \der(\fA)$. Such connections will be called antihermitian connections. Then any unitary element $U\in \fA$ with $\det(U) = \gone$ defines on $\fA$ a right module endomorphism $S \mapsto US$ which preserves the hermitian structure and the $\det$ application. We denote by $SU(\fA)$ the group of such elements of $\fA$. In our particular case, this is exactly the gauge group of the $SU(n)$-vector bundle $\cE$. We denote by $U(\fA)$ the group of unitary elements of $\fA$. For any $U\in U(\fA)$, the gauge transformation of a noncommutative connection $\widehat{\nabla}$ is defined by the relation $\widehat{\nabla}^U_\cX S = U^\ast \widehat{\nabla}_\cX (US)$. The noncommutative $1$-form $\omega$ is then transformed as
\begin{align*}
  \omega \mapsto U^\ast \omega U + U^\ast \hd U
\end{align*}

Any ordinary connection on $\cE$ defines canonically a noncommutative connection on $\fA$. Indeed, such a connection is given by a noncommutative $1$-form $\alpha$. One then defines a noncommutative connection $\widehat{\nabla}^\alpha$ by 
\begin{align*}
  \widehat{\nabla}_\cX^\alpha S := \nabla_X S + S \alpha(\cX) = \cX S + \alpha(\cX)S
\end{align*}
for any $\cX \in \der(\fA)$ and $S\in \fA$. The curvature of this connection coincides with the ordinary curvature $\hat{R}^\alpha(\cX, \cY) = R^\cE(X,Y)$ and this noncommutative connection $\widehat{\nabla}^\alpha$ is compatible with the hermitian structure on $\fA$. Finally, a gauge transformation on $\nabla^\cE$ induces a $SU(\fA)$-gauge transformation on $\widehat{\nabla}^\alpha$.

This means that noncommutative connections on $\fA$ are extensions of ordinary connections on $\cE$. In~\cite{dubois-violette:90:II,dubois-violette:89,dubois-violette:89:II,dubois-violette:98}, it was shown that the extra degrees of freedom can be interpreted as Higgs fields. We refer to these papers for details.

Because $\fA$ and  $C^{\infty}(M)$ are Morita equivalent, their projective right modules are in bijection (the $K$-groups are the same). From the physical point of view this means that the matter contents of any associated theory does not permit one to distinguish between $C^{\infty}(M)$ and $\fA$.

\subsection{Symmetries and noncommutative connections}

In the following section, we will be interested in noncommutative connections invariant under the action of a Lie group $G$. Here we give a precise definition of this concept.

Let $\cG$ be the Lie algebra of $G$. An action of $\cG$ on $\fA$ is a Cartan operation of $\cG$ on the graded differential algebra $\Omega_\der(\fA)$. In particular, any element $\cG$ can be considered as an element in $\der(\fA)$, which means that we will always look at $\cG$ as a Lie subalgebra of $\der(\fA)$.

A $\cG$-invariant noncommutative connection on the right module $\fA$ is a noncommutative connection  $\widehat{\nabla}$ satisfying
\begin{align*}
  Y\left( \widehat{\nabla}_\cX a \right) = \widehat{\nabla}_{[Y,\cX]} a +
  \widehat{\nabla}_\cX \left( Ya \right)
\end{align*}
for any $Y\in \cG$, $\cX \in \der(\fA)$ and $a\in \fA$. If $\widehat{\nabla}$ is given by the noncommutative $1$-form $\omega$, this is equivalent to
\begin{align*}
  L_Y \omega = 0
\end{align*}
for any $Y\in\cG$.

 \subsection{$\Omega_\der(\fA)$ and $\Omega(E)$}\label{sec:omega_der}

In the next section, we will characterize the $\cG$-invariant connections on the right module $\fA$, where $\fA$ is the algebra of endomorphisms of a $SU(n)$-vector bundle $\cE$. In order to do that, it is very convenient to look at $\Omega_\der(\fA)$ in a different way, using a result proved in~\cite{masson:99}. There, $\Omega_\der(\fA)$ was shown to be some basic subalgebra of a bigger differential graded algebra. Let us describe this algebra and give new results about this very useful construction.

Let us denote by $E$ the principal $SU(n)$-bundle over $M$ for which $\cE$ is associated, and denote by $C^\infty(E)$ the (commutative) algebra of smooth functions on $E$. Then one has a map $\xi \mapsto \xi^E$ which sends any $\xi \in \tsu(n)$ into the associated vertical vector field on $E$. Let us introduce the algebra $\fB = C^\infty(E) \otimes M_n$ of matrix valued functions on $E$.   Denote by $(\Omega_\der(\fB), \hd) = (\Omega(E) \otimes \Omega_\der(M_n), \d + \d')$ its differential calculus based on derivations. It is easy to see that $\{ \xi^E + ad_\xi\ / \ \xi \in \tsu(n) \}$ is a Lie subalgebra of $\der(\fB)$ which is isomorphic to $\tsu(n)$. This Lie subalgebra defines a Cartan operation of $\tsu(n)$ on $\Omega_\der(\fB)$, whose basic subalgebra we denote by $\Omega_{\der,\Bas} (\fB)$. Then it was proved in~\cite{masson:99} that $\Omega_\der(\fA) = \Omega_{\der,\Bas} (\fB)$
 
A $SU(n)$-connection on $E$ is given by a $1$-form $\omega_E$ on $E$ with values in $\tsu(n) \subset M_n$. This connection defines a connection on $\cE$ (denoted by $\nabla$), which itself gives rise to a noncommutative $1$-form $\alpha \in \Omega^1_\der(\fA)$. From the previous result, this form comes from a basic $1$-form $\alpha^E$ in $\Omega_{\der,\Bas} (\fB)$, which is nothing but $\alpha^E= \omega_E - i \theta$, where $\theta \in \Omega^1_\der(M_n)$ is the canonical $1$-form defined previously. The basicity of this $1$-form is a consequence of properties of $\omega_E$ and $i\theta$, in particular the equivariance of $\omega_E$. 

At the level of derivations, the relations between the algebras $\fA$ and $\fB$ can be summarized in the following exact commutative diagram which combines derivations on $\fA$, derivations on $\fB$ and vector fields on $E$:
\begin{equation}
\label{diagderivations}
\raisebox{.5\depth}{%
  \xymatrix{
    &           &  0 \ar[d]       &     0  \ar[d]         &    \\
    &  0    \ar[d]\ar[r]    & \cZ_{\der}(\fA) \ar[d]\ar[r]   &  \G(TVE)  \ar[d]\ar[r]   & 0  \\
    0 \ar[r] & \int(\fA) \ar[d]\ar[r]  & \cN_{\der}(\fA) \ar[d]_{\tau}\ar[r]^{\rho_E}   &  \G_{M}(E)  \ar[d]^{\pi_*}\ar[r] & 0  \\
    0 \ar[r]& \int(\fA) \ar[d]\ar[r]  &\der(\fA) \ar[d]\ar[r]_{\rho}   & \G(TM)  \ar[d]\ar[r]      & 0  \\
    &  0        &     0     &    0           &  
    }}
\end{equation}
The lower row is just the ordinary short exact sequence which relates vector fields on $M$, derivations on $\fA$ and inner derivations on $\fA$. In the middle column, $\cN_{\der}(\fA) \subset \der(\fB)$ is the subset of derivations on $\fB$ which preserve the basic subalgebra $\fA \subset \fB$, and $\cZ_{\der}(\fA) \subset \der(\fB)$ is the subset of derivations on $\fB$ which vanishes on $\fA$. These two Lie algebras were defined for more general algebras in \cite{masson:96}. The short exact sequence they define is the one used to prove that $\fA$ is a  noncommutative quotient manifold of the noncommutative algebra $\fB$~\cite{masson:99}. The Lie algebra $\cZ_{\der}(\fA)$ is generated as a $C^\infty(E)$-module by the particular elements $\xi^E + ad_\xi$ for any $\xi \in \tsu(n)$. 

The right most column involves only geometrical objects. The space $\G_{M}(E)$ is defined to be
\begin{align}
  \G_{M}(E)= \{ \hX \in \G(E) / \pi_* \hX(p) = \pi_*\hX(p') \ \forall p , p' \in E \text{ such that }  \pi(p)=\pi(p')     \}
\end{align}
This is the Lie algebra of vector fields on $E$ which can be mapped to vector fields on $M$ using the tangent maps $\pi_* : T_pE \rightarrow T_{\pi(p)}M$.

In the following, for any $\eta = f \otimes \xi \in C^\infty(E) \otimes \tsu(n)$, we will denote by $\eta^E \in \G(TVE)$ the vector field $f  \xi^E$ on $E$. Using this notation, any element in $\cZ_{\der}(\fA)$ can be written as $\eta^E + ad_\eta$ where $\eta \in C^\infty(E) \otimes \tsu(n)$. The isomorphism between $\cZ_{\der}(\fA)$ and $\G(TVE)$ maps any element $\eta^E + ad_\eta$ into $\eta^E$. 

The diagram (\ref{diagderivations}) bears some strange similarities with the diagram presented on page 12 in \cite{dubois-violette:98} which involved Lie algebroid structures. We will not make further comments about this point here.

A connection $\omega_E$ on $E$ splits three short exact sequences in this diagram, in a compatible way. First, this connection can be used to lift vector fields $X$ on $M$ into horizontal vector fields $X^h$ on $E$. This gives us a splitting map of 
\begin{equation}
\xymatrix@1{0 \ar[r] & \G(TVE) \ar[r] & \G_{M}(E) \ar[r] & \G(TM) \ar[r] & 0}
\end{equation}
as $C^\infty(M)$-modules. It was shown in \cite{dubois-violette:98}, and recalled in \ref{algebra-endom-vect}, that the connection $\omega_E$ splits the short exact sequence (\ref{sesder}) of $C^\infty(M)$-modules using the map $X \mapsto \nabla_X$. Now, using notations introduced so far, there is a splitting of the middle column by the map $\cX \mapsto \cX^E= \rho(\cX)^h - ad_{\alpha(\cX)^E}$ where $\alpha(\cX)^E$ is the basic element in $\fB$ associated to $\alpha(\cX) \in \fA$ (notice that $\alpha(\cX)^E=\alpha^E(\cX^E)$).

These splittings can be used to decompose any element in $\cN_{\der}(\fA)$ into four parts, making explicit the kernels of the two short exact sequences in which this space is involved. Any derivation $\tX \in \cN_{\der}(\fA)$ can be written, as a derivation on $\fB$, in the form $\tX= \hat{X} + ad_\gamma$, where $\hat{X} \in \G(E)$ and $\gamma \in C^\infty(E)\otimes \tsl_n$. At this stage, it is easy to directly show that $\hat{X}$ is in $\G_M(E)$ using the restriction of $\tX$ on the center $C^\infty(M)$ of $\fA$ considered as a subalgebra of $\fB$. A derivation $\tX$ belongs to $\cN_{\der}(\fA)$ if and only if  $L_\xi \tX \in \cZ_{\der}(\fA)$ for any $\xi \in \tsu(n)$. This means that there must exist $\eta \in C^\infty(E) \otimes \tsu(n)$ such that:
\begin{align}
[\xi^E, \hat{X}] &= \eta^E \\
L_\xi \gamma &= \eta
\end{align}
Applying a connection $\omega_E$ on the first relation, and using the equivariance of $\omega_E$, one gets $L_\xi (\omega_E(\hat{X})) = \eta$. Let us introduce $Z =- \alpha^E(\tX)= \gamma - \omega_E(\hat{X}) \in C^\infty(E) \otimes \tsl_n$. Then $L_\xi Z = 0$ for any $\xi \in \tsu(n)$, which implies that $Z \in \fA_0$, or $ad_Z \in \int(\fA)$. Using this result, the derivation $\tX$ can be written as $\tX = \hat{X} + ad_\gamma = X^h + \hat{X}^v + ad_{\omega_E(\hat{X}) + Z}$. With our notations, one has $\hat{X}^v = \omega_E(\hat{X})^E$ (vertical part of the vector field $\hat{X}$). So, one can write finally
\begin{equation}
  \tX = X^h +ad_Z + \underbrace{\omega_E(\hat{X})^E + ad_{\omega_E(\hat{X})}}_{ \in \cZ_{\der}(\fA)} = X^h + \omega_E(\hat{X})^E  + ad_{\omega_E(\hat{X})} +  \underbrace{ad_{Z}}_{ \in \int(\fA)} 
\end{equation}
The situation can be summarized in the following diagram where all the splittings are explicit
\begin{equation}
  \label{eq:diagramme}
  \raisebox{.5\depth-.5\height}{%
    \xymatrix@R=0pc@C=0pc@M=2pt{
      \cN_{\der}(\fA) \ar@{->>}[rrrrrrr]^{\rho_E} \ar@{->>}[dddddd]_{\tau} &&&& \rule{60pt}{0pt} &&& \G_M(E)
      \ar@{->>}[dddddd]^{\pi_*} \\
      &&&\makebox[100pt][r]{$(\pi_*\hX)^h+\omega_E(\hX)^E+\ad_{\omega_E(\hX)}$}& & \hX \ar@{|->}[ll]&&\\
      &\makebox[15pt][l]{$\cX^E = \rho(\cX)^h-\ad_{\alpha(\cX)}$} &&&&& X^h & \\
      &&&&\rule{0pt}{30pt}&&&\\
      &\cX \ar@{|->}[uu] &&&&& X \ar@{|->}[uu] & \\
      &&\nabla_X &&& X \ar@{|->}[lll]&&\\
      \der(\fA) \ar@{->>}[rrrrrrr]_{\rho} &&&&&&& \G(TM)
      }}
\end{equation}

In the following $\Omega_\der(\fA)$ will be identified with the corresponding basic subalgebra of $\Omega_{\der}(\fB)$. Let us now look at the consequences of this construction on a noncommutative  connection given by a $1$-form $\omega \in \Omega^1_\der(\fA)$. Such a 1-form can be decomposed as
\begin{align*}
  \omega = a - \phi \in [ \Omega^1(E) \otimes M_n ] \oplus [ C^\infty(E)   \otimes M_n \otimes \tsl_n^\ast ]
\end{align*}
with the basic conditions 
\begin{gather*}
(L_{\xi^E} + L_{ad_\xi})a = 0  \ \ \ \ \ \ \ \  (L_{\xi^E} + L_{ad_\xi})\phi = 0 \\
i_{\xi^E} a - i_{ad_\xi}\phi = 0
\end{gather*}
for any $\xi \in \tsu(n)$. Here we use obvious notations for geometrical and algebraic parts of the Lie derivative $L$ and the interior product $i$.
 
At this level, some general comments  about these relations are in order. First, the invariant relation on $a$ is nothing but the covariance relation of an ordinary connection on $E$. But the last relation prevents $a$ to be such a connection. Indeed, this last relation generalizes the vertical condition on ordinary connections, and connect the value of $a$ on vertical vector fields to the values of $\phi$. For ordinary connections on $E$, $\phi$ being replaced by $i \theta$, the usual vertical condition on $a$ is recovered. The second relation, the invariance of $\phi$, has a natural geometric interpretation. By its very definition, $\phi$ can be viewed as a map $E \rightarrow M_n \otimes \tsl_n^\ast$. Using standard results in differential geometry, the invariance relation on $\phi$ permits one to interpret $\phi$ as the section of a vector bundle over $M$, associated to $E$, whose fiber is $M_n \otimes \tsl_n^\ast$. In this identification, the Lie derivative $L_{ad_\xi}$ on $M_n \otimes \tsl_n^\ast$ is nothing but the infinitesimal action of the Lie group $H$ on $M_n \otimes \tsl_n^\ast$ involved in the construction of this associated vector bundle. We will use such an identification in similar cases several times in the following.

Let us now consider the following situation, which will be the starting point for the next section and which generalizes to the noncommutative connections what has been presented in section~\ref{construction_1} on ordinary connections. Assume we have an action of a compact connected Lie group $G$ on the principal fiber bundle $E$ which commutes with the natural right action of the Lie group $H=SU(n)$ on $E$. Then for any $Y\in \cG$, the Lie algebra of $G$, one can associate a vector field $Y^E$ on $E$. This vector field induces a Cartan operation of $\cG$ on $\Omega(E)$. This operation extends naturally to an operation on $\Omega_\der(\fB) = \Omega(E) \otimes \Omega_\der(M_n)$ where $\cG$ acts only on the $E$ part. Because the actions of $G$ and $H$ commute, the operation of $\cG$ respects the basic subalgebra $\fA$ of $\fB$, and restricts to an operation on $\Omega_\der(\fA)$. Then the original action of $G$ on $E$ gives rise to a (noncommutative) action of $\cG$ on $\fA$. This action is the one we will use to characterize $G$-invariant noncommutative connection on $\fA$.

\subsection{Local point of view}
In this section, we want to study more precisely the relation between the algebras $\fA$ and  $\fB$ from the local point of view.

Let us first characterize local objects in $\fA$. Such a discussion was preformed in \cite{dubois-violette:98,masson:99} and we recall essential points here. Over an  open subset $U$ over which the fiber bundle $\End(\cE)$ is trivialized, the algebra  $\fA$ is isomorphic to $\fA\loc:=C^{\infty}(U)\otimes M_n$, and one can associate to any element $a\in \fA$ an element $a\loc\in \fA\loc$. Over an intersection $U\cap U' \neq \emptyset$ of two such open sets $U$ and $U'$, one has a transition function $g: U\cap U' \to SU(n)$ which relates $a\loc'$ to $a\loc$ in the following way
\begin{align*}
  a\loc'=\Ad_{g^{-1}} a\loc
\end{align*}

One can also associate to any derivation $\cX \in \der(\fA)$ a local derivation $\cX\loc \in \der(C^{\infty}(U)\otimes M_n)$. Such a derivation can be decomposed into two parts $\cX\loc=X_{|U} +\ad_{\gamma\loc}$, where $X=\rho(\cX)$ (see equation~(\ref{rho})), and $X_{|U}$ is its restriction to the open subset $U$. It is possible to give an explicit expression for $\gamma\loc$ if one consider a connection on $\End(\cE)$, to which one can associate the noncommutative 1-form $\alpha$ and the local 1-form $A\loc \in \Omega^1(U)\otimes \cH$.
Then one has
\begin{align*}
  \gamma\loc&= A\loc(X_{|U})-\alpha(\cX)\loc
\end{align*}
Over an intersection $U\cap U' \neq \emptyset$,  $\cX\loc'$ and  $\cX\loc$ are related in the following way
\begin{align*}
  X_{|U}' &=X_{|U}\\
  \gamma\loc' &= \Ad_{g^{-1}}\gamma\loc + g^{-1} X_{|U}(g)
\end{align*}

Finally, let us consider local 1-forms, and associate to an element $\omega \in \Omega_{\der}^1(\fA)$ over $U$, an element $\omega\loc \in \Omega_{\der}^1(C^{\infty}(U)\otimes M_n)$. Such an element can be decomposed into two parts: $\omega\loc= a + \phi\circ i\theta$ where $a \in \Omega^1(U)\otimes M_n$, and $\phi \in C^{\infty}(U)\otimes M_n \otimes \tsl_n^*$.
Then over an intersection $U\cap U'\neq \emptyset$,  $\omega\loc$ and  $\omega\loc'$ are related in the following way:
\begin{equation}
  \begin{split}
    a' &= \Ad_{g^{-1}} \circ a - \Ad_{g^{-1}} \circ \phi \circ \Ad_{g} \circ g^*\theta^H \\
    \phi' &= \Ad_{g^{-1}} \circ \phi \circ \Ad_{g}
  \end{split}
  \label{transition-relation}
\end{equation}
where $\theta^H$ is the usual Cartan form on the group $H$, and $g^*\theta^H= g^{-1}dg$. One can remark that this transition relations look like the one encountered in usual commutative gauge theories, but in the present case these relations are ``twisted'' by the scalar field $\phi$. It was shown in \cite{masson:99} that if one choose a reference connection, then one can express local forms in term of tensors which transform in a much more manageable way.

Let us now make some remarks about the relations between  these local objects and the local ones associated to the algebra $\fB = C^{\infty}(E)\otimes M_n$. First, let us consider a local section $s: U \to E$. One can associate to it a trivial extension of the pullback application $s^*: \fB \to C^{\infty}(U)\otimes M_n$. Then by definition, the image of $\fB_{\cH-\bas |U}$ by the application $s^*$ is the algebra $\fA\loc$ obtained from the localization of $\fA$ over $U$.

For derivations, it was shown previously that with the help of an ordinary connection, one can associate to any element $\cX \in \der(\fA)$ an element $\cX^E \in \cN(\fA) \subset \der(\fB)$, where explicitely  $\cX^E= \rho(\cX)^h - \ad_{\alpha(\cX)^E}$. In a similar way, using the inclusion $ \Omega_{\der}(\fA) \hookrightarrow \Omega_{\der}(\fB)$, one can associate to any element $\omega \in \Omega_{\der}(\fA)$ an element\footnote{In opposition to the rest of this paper, we use here explicit different notations $\omega$ and $\varpi$.} $\varpi \in \Omega_{\der}(\fB)_{\cH-\bas}$. Then one can compare the expressions $\varpi(\cX^E)$ and $\omega(\cX)$. Over  an open subset $U \subset M$ over which the principal fiber bundle $E$ is trivialized, using the local expression of the connection, one has
\begin{equation*}
  \cX^E_{|s(x)} = s_* X_{|x}- A\loc(X)_{|x}- \ad_{\alpha(\cX)^E_{|s(x)}}
\end{equation*}
where $x \in U$. From the basicity of the 1-form $\varpi$, one can then show  that $\omega\loc(\cX\loc)=s^*(\varpi_{|U}(\cX^E_{|U})) = s^*\varpi_{|U}(\cX\loc)$, so that $s^*\varpi_{|U}=\omega\loc$. This result generalizes the previous result about elements of the algebras, and one has $s^*\Omega_{\der}(\fB)_{\cH-\bas |U}=\Omega_{\der}(\fA\loc)$.  This relation shows that one can obtain the local expression $\omega\loc$ either from the 1-form  $\omega \in \Omega_{\der}^1(\fA)$, or from the 1-form $\varpi \in \Omega_{\der}^1(\fB)_{\cH-\bas}$. Finally, notice that the transition relations~(\ref{transition-relation}) could have been obtained   by considering the 1-form $\varpi$ over the intersection  $U\cap U' \neq \emptyset$ of two open sets, with the transition relation $s'= s \cdot g$ between local sections $s: U\to E$ and $s':U'\to E$.


\section{Invariant noncommutative connections} \label{invar-nonc-conn} 

Here, we characterize the degrees of freedom of invariant noncommutative connections in the setting exposed in the previous section. The results obtained are generalizations of the results summarized in section~\ref{construction_1} for ordinary invariant connections. In particular, the constructions presented in section~\ref{construction_1} are explicitely used in the present case. Indeed, it is possible to make reference to the structure of diagram~(\ref{diagram}) thanks to the trick exposed at the end of the previous section, which consists to look at $\fA$ as the basic subalgebra of $\fB= C^\infty (E) \otimes M_n$ for a well chosen Cartan operation of $\cH=\tsu(n) \subset M_n$.
The starting ingredients of this section are the following. A $G$-invariant noncommutative connection $\omega \in (\Omega^1_{\der}(\fA))_{\cG-\inv}$ is written  as a basic element  
\begin{align*}
\omega = a - \phi \in [\Omega^1(E) \otimes M_n] \oplus [ C^{\infty}\otimes M_n \otimes \der(M_n)^*]
\end{align*}
Then the objects $a$ and $\phi$ satisfy the 3 relations
\begin{align}
  L_{\xi} ( a-\phi) &= 0 \label{invh} \\
  i_\xi( a-\phi) &= 0 \label{hor} \\
  L_{X} ( a-\phi) &= 0 \label{invg} 
\end{align}
for any $\xi \in \cH=\tsu(n)$ and any $X \in \cG$. Recall that the ordinary connections are those for which $\phi=i\theta$, which is a straightforward way to recover all the results of section~\ref{construction_1} from the results presented here.

\subsection{Global approach}\label{global_approach}

This approach is similar to the  one performed in section~\ref{construction_1}, and it uses essentially the same technics. With the help of the $G$-invariance conditions~(\ref{invg}), we can restrict the dependence of  $a$ and $\phi$ to $Q \subset E$. Then $a$ is completely determined from its values on $T_qE$ for all $q\in Q$. The application $a_q : T_q E \to M_n$  can be  decomposed into several parts. Let us call $\mu \in \Omega^1(Q)\otimes M_n$ the restriction of the 1-form $a$ to the tangent space $TQ$. By relation~(\ref{hor}), one has  $\mu_q(\xi_q^E) = \phi_q(\xi)$ for  any $\xi \in \zo$. This 1-form satisfies the equivariance property
\begin{align*}
  L_{(X,\xi)}\mu &= ( L_{X} +L_{\xi} )\mu =0   &
 \forall  (X,\xi)& \in \cN_{\so} \subset \cG\times\cH \ .
\end{align*}
(there, we use notations of section~\ref{inv_con}. Recall also that $L_{\xi}$ contains a geometric and an algebraic part)\\
Then, using $\so$-invariance on $Q$, it is easy to show that $\mu$ takes its value in the vector space 
\begin{align*}
  \wo & := Z( \lambda_* \cG_0, M_n) \ ,
\end{align*}
 the centralizer of $  \lambda_*  \cG_0$ in $M_n$. As a trivial consequence, $\eta_q :=\phi_{q|\zo}$ has also its values in $\wo$. 

A simple analysis shows that $\wo$ is an associative subalgebra of $M_n$ on which the Lie algebra $\zo$ acts by  the adjoint action. It is natural and useful for the following to associate to it the differential calculus $\Omega_{\zo}(\wo)=\wo \otimes \bigwedge\zo^*$ which mimics the differential calculus $\Omega_{\der}(M_n)=M_n \otimes \bigwedge\tsl_n^*$. On $\Omega_{\zo}(\wo)$, the differential is defined as in formula~(\ref{differential}), where now the Lie algebra $\zo$ plays the role of the derivations on the algebra $\wo$. An other important useful feature of $\wo$ is that there is a natural application
\begin{align*}
  \cN_{ \so} &\to \der\left( C^{\infty}(Q) \otimes  \wo \right)\\
  (X,\xi) &\mapsto  X^Q+\xi^Q + ad_{\xi}
\end{align*}
Notice that $\so$ is sent to zero in this application. This implies that this application factorizes through an application $\cN_{\so}/\so \to \der\left( C^{\infty}(Q) \otimes  \wo \right)$, which permits us to define a Cartan operation of $\cN_{\so}/\so$ on $\Omega(Q)\otimes \Omega_{\zo} (\wo)$, whose Lie derivation is denoted by 
\begin{align*}
  L_{(X,\xi)} & = L_{X^Q+\xi^Q} + L_{ad_{\xi}} &
  &\forall (X,\xi) \in \cN_{\so}/\so=\cK\oplus\zo
\end{align*}
In particular, this operation induces an operation on the Lie algebra $\zo$, denoted by $L_\xi$, for any $\xi \in \zo$.

The difference $\mu-\eta$ is naturally an element of degree 1 in $\Omega(Q)\otimes \Omega_{\zo}(\wo)$. Using relations~(\ref{invh}),~(\ref{hor})~and~(\ref{invg}), it is easy to verify that 
\begin{align*}
  i_{\xi} (\mu -\eta) &= 0 &
  &\forall \xi \in \zo \\
  L_{(X,\xi)} (\mu -\eta) &= 0 &
  &\forall  ( X,\xi) \in \cN_{\so}
\end{align*}
This implies that $\mu -\eta \in \left( \Omega(Q) \otimes \Omega_{\zo}(\wo)\right)_{\zo-\bas}^1$.

Now, let us introduce $\alpha\in C^{\infty}(Q)\otimes \wo\otimes \bigwedge \cK^*$, defined by
\begin{align*}
  \alpha(X) &= \mu(X^Q) & 
  & \forall X\in \cK
\end{align*}
One can show that $\alpha \in \wo \otimes ( C^{\infty}(Q) \otimes \bigwedge \cK^*)_{\cK-\inv}$, where the $\cK$-invariance is defined by the induced Lie derivative of the operation of $\cN_{\so}$ on the $C^\infty(Q)$ part and by the standard Lie derivative on the $\bigwedge \cK^*$ part (the differential and the Lie derivative on $\bigwedge \cK^*$ are naturally induced by the Lie algebra structure of $\cK$).

Let us now introduce the  bigger differential calculus 
\begin{align*}
  \Omega_{\zo +\cK}(M/G,\wo) := ( \Omega(Q) \otimes \Omega_{\zo} (\wo) \otimes   \bigwedge \cK^*)_{(\zo+\cK)-\bas}
\end{align*}
equipped with  the natural differential which is the sum of the differentials on each components. Latter, we will make some comments about this differential algebra, in particular we will explain why $M/G$ makes its appearance in the notation.

The main result of the previous discussion is that we can show that  $\mu -\eta-\alpha \in \Omega^1_{\zo+\cK}(M/G,\wo)$. This relation permits one to characterize in a common algebraic 1-form the restrictions of $a$ and $\phi$ to $TQ$. 

Let us now look at the other parts of the space $T_qE$, that is  $\cL^Q_{|q}$ and  $\cM^Q_{|q}$. Using similar arguments as in \ref{inv_con}, the restriction $ \psi := a_{|\cL^Q}$ defines a section of the vector bundle  associated to \fiber{N(\so)/\so}{Q}{M/G}  whose fiber is the vector space
\begin{align*}
  \cF_{\cL}&:=  ( M_n  \otimes \cL^*)_{\so-\inv} =\{\ell: \cL \to M_n / L^{\cL}_{(X,\lambda_* X)} \ell =0 \ \ \forall X \in \cG_0\}
\end{align*}
where $ (L^{\cL}_{(X, \xi)}\ell)(Y)= - \ell([X,Y])+ [\xi,\ell(Y)]$ for any $(X,\xi) \in N(\so)$. This relation is the natural action of $\cN_{\so}$ on the space $M_n \otimes \cL^*$. For this action, $\cF_\cL \subset M_n \otimes \cL^*$ is invariant. The fact that $\psi \in C^\infty(Q)\otimes \cF_\cL$ is a section of a vector bundle comes from the equivariance property
\begin{align*}
  L_{(X,\xi)} \psi&= (L_{X^E+\xi^E}+ L^{\cL}_{(X,\xi)})\psi=0 &
  &\forall (X,\xi) \in \cN_{\so}
\end{align*}
  Therefore, one has $\psi \in \left(C^{\infty}(Q) \otimes   \cF_{\cL}  \right)_{(\zo+\cK)-\inv} $.

  In the same way, the restriction $\zeta := a_{|\cM^Q}=\phi_{|\cM}$, considered as an element in $C^\infty(Q) \otimes M_n\otimes\cM^*$, is a section of the vector bundle  associated to \fiber{N(\so)/\so}{Q}{M/G} whose fiber is the vector space 
  \begin{align*}
    \cF_{\cM} &:=  ( M_n  \otimes \cM^*)_{\so-\inv} =\{m: \cM \to M_n / L^{\cM}_{(X , \lambda_* X)} m =0 \ \ \forall X \in \cG_0\}
  \end{align*}
where $ (L^{\cM}_{(X,\xi)}m)(Y)= - m([X,Y])+ [\xi,m(Y)]$ for any $(X,\xi) \in N(\so)$. As before, $\zeta$ satisfies the equivariance property
\begin{align*}
L_{(X,\xi)} \zeta&= (L_{X^E+\xi^E}+ L^{\cM}_{(X,\xi)})\zeta=0 &
&\forall  (X,\xi) \in \cN_{\so}
\end{align*}  
and so $\zeta \in \left(C^{\infty}(Q) \otimes   \cF_{\cM}  \right)_{(\zo+\cK)-\inv} $
  
Now, noticing that 
\begin{align*}
  ( M_n  \otimes \cL^*)_{\so-\inv} \oplus  (  M_n \otimes \cM^* )_{\so-\inv} &=  ( M_n \otimes (\cL^* \oplus \cM^* )_{\so-\inv}=:\cF
\end{align*}
$\zeta +\psi$ can be considered as a section of the associated vector bundle to \fiber{N(S_0)/S_0}{Q}{M/G} where the fiber is the bigger vector space $\cF$.
Collecting all the previous degrees of freedom, we have proven that
\begin{align*}
  (\Omega^1_{\der}(\tA))_{\cG-\inv} \simeq \Omega^1_{\zo +\cK}(M/G,\wo) \oplus \cP
\end{align*}
where
\begin{align*}
  \cP &= (C^{\infty}(Q) \otimes \cF)_{(\zo+\cK)-\inv}
\end{align*}
Denote by $\cC$ the algebra $ (C^{\infty}(Q) \otimes \wo)_{(\zo+\cK)-\inv}$. Then $\Omega_{\zo+\cK}(M/G,\wo)$ is  a differential calculus associated to $\cC$ in the sense that  $\Omega^0_{\zo+\cK}(M/G,\wo)=\cC$. This algebra can be interpreted as the sections of a fiber bundle associated to  \fiber{N(S_0)/S_0}{Q}{M/G} whose fibers are modeled over the algebra $\wo$. The algebra $\cC$ can be considered as a ``reduction'' of the algebra $\fA$. As a matter of fact, it is easy to verify that the elements in $SU(\cC)$ define noncommutative gauge transformations on the space of $\cG$-invariant connections on $\fA$. Equipped with the differential calculus $\Omega_{\zo+\cK}(M/G,\wo)$ and the module $\cP$, this algebra is the natural building block, in a noncommutative viewpoint, for the $\cG$-invariant connections on $\fA$. 

As a last remark, notice that all the objects introduced here are naturally related to fiber bundles over the reduced space $M/G$, as it was also the case for ordinary $\cG$-invariant connections. As a matter of fact, the situation is rather similar to the classical one, but here, there are new scalar fields coming from noncommutative geometry in addition to those coming from dimensionnal reduction. We can also notice that in the noncommutative framework, we do not need a reference connection to obtain objects which ``live'' over $M/G$.

\subsection{Local approach}\label{local_approach}

As in the classical case, we will study invariant connections in the case where $M=M/G\times~G/G_0$. We will use the notations introduced in section~\ref{wang_approach}. The idea developed in section~\ref{wang_approach} is to pull back  an invariant connection on $\tE$ to a bigger space $Q'\times H$. There, an invariant connection can be written in a very compact and elegant generic form using in particular the Cartan 1-forms on the groups $G$ and $H$. Here we generalize this construction. In order to do that, we need to find the ``good'' space on which pulling back the invariant noncommutative connection. Because we work in a noncommutative framework, we have to deal with algebras instead of spaces. The following diagram summarizes the relations between the spaces and algebras we consider
\begin{align*}
\raisebox{.5\depth-.5\height}{%
  \xymatrix{
    Q'\times H \ar[d]_{\Psi} &{\xy *[o]=<30pt>\hbox{?}="o"*\frm{o} \endxy}  \ar@{^(->}[r] & \Omega(Q'\times H)\otimes \Omega_{\der}(M_n) \\
    \tE \ar@{~>}[r] & \Omega_{\der}(\fA) \ar@{^(->}[r]^-{\text{basic}}_-{\text{for $\cH$}} & \Omega(\tE)\otimes \Omega_{\der}(M_n) \ar[u]^{\Psi^*}   
    }}
\end{align*}

The algebra we are  looking at must replace the algebra $\fA$, in the same way the space $Q'\times H$ replaces  the space $\tE$. It is natural to look at this algebra as a basic subalgebra of $C^{\infty}(Q'\times H)\otimes M_n$ for the Cartan operation of $\cH$. Then the pull back $\pso$ of a $\cG$-invariant connection 1-form $\omega \in \Omega_{\der}(\fA) \subset \Omega(\tE)\otimes \Omega_{\der}(M_n) $ belongs to $ \Omega(Q'\times H)\otimes \Omega_{\der}(M_n)$. Thanks to the facts that $Z_0$ and $G_0$ do not act on the $M_n$ part of these algebras, that $\Psi^* $ preserves the $\cG$-invariance and the $\cH$-basicity, one has
\begin{align*}
  \pso \in \left[ \Omega(\tQ) \otimes \Omega(G) \otimes \Omega(H) \otimes \Omega_\der(M_n) \right]_{\substack{G-\inv \hfill \\ H-\bas \hfill\\(Z_0\times G_0)-\bas \hfill}} \ .
\end{align*}

The advantage to work at the level of the space $\Omega(\tQ) \otimes \Omega(G) \otimes \Omega(H) \otimes \Omega_{\der}(M_n)$, is that, there, we can decompose in an easy way the actions of the group $G$, $G_0$, $Z_0$ and $H$. These actions are shown in the following suggestive diagram
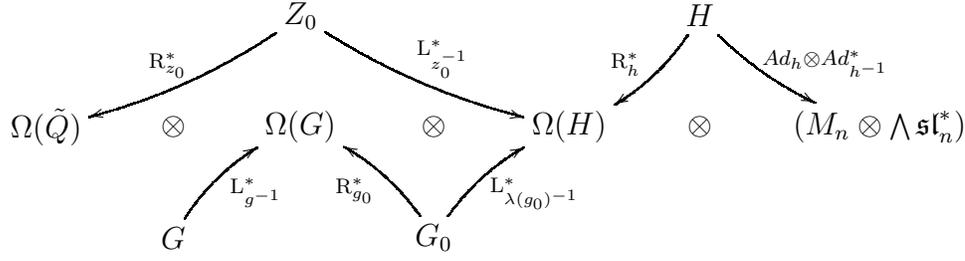
\begin{align*}
  \xymatrix{
    &&Z_0 \ar@/^/[dll]_{\R^*_{z_0}} \ar@/_/[drr]^{\L^*_{z_0^{-1}}}&&&H \ar@/^/[dl]_{\R^*_{h}} \ar@/_/[dr]^{Ad_{h}\otimes Ad^*_{h^{-1}} }&\\
    \Omega({\tQ}) & \otimes & \Omega(G) & \otimes& \Omega(H) & \otimes&  \left(M_n\otimes \bigwedge \tsl_n^*\right) & \\
    &G \ar@/^/[ur]_{\L^*_{g^{-1}}}&& G_0 \ar@/_/[ul]^{\R^*_{g_0}} \ar@/^/[ur]_{\L^*_{\lambda(g_0)^{-1}}}&&&
    }
\end{align*}
where symbols $\L$ and $\R$ mean respectively left and right multiplication. Notice that the actions  of $Z_0$ and $G_0$ on $\Omega(H)$ commute.

Then, using $H$-basicity and $G$-invariance, a straightforward computation shows that $\pso$ can be written in the generic form
\begin{align}
  \pso_{|(\tq,g,h)} = Ad_{h^{-1}} \left( \tom_{|\tq} + \Lambda_{\tq}\circ\theta^G_{|g} +  \phi_{\tq}\circ Ad_{h}\circ (\theta^H_{|h}-i\theta)\right)\label{decomposition}
\end{align}
where $\theta^G$ and $\theta^H$ are the Cartan 1-forms on the groups $G$ and $H$ and $i\theta$ is the algebraic 1-form introduced in section~\ref{ncdc}. It is very natural to use the 1-form $i\theta$ in this relation in order to make explicit the identification~(\ref{soudure}). Notice that $\theta^H$ and $i \theta$ are known to look very similar in their structure~\cite{dubois-violette:90}, and that they  appear together in this very compact expression.
In formula~(\ref{decomposition}), on has
 \begin{align*}
   \tom &\in \Omega^1(\tQ) &
   \Lambda &\in C^{\infty}(\tQ)\otimes M_n \otimes \cG^*  &
   \phi & \in  C^{\infty}(\tQ)\otimes M_n \otimes \tsl_n^*
 \end{align*}
 and the $Z_0$-invariance implies that
 \begin{align*}
   \R^*_{z_0} \tom  &= Ad_{z_0^{-1}} \circ \tom &
   \R^*_{z_0} \Lambda &= Ad_{z_0^{-1}} \circ \Lambda &
   \R^*_{z_0} \phi &= Ad_{z_0^{-1}} \circ \phi \circ Ad_{z_0} 
 \end{align*}
for any $z_0 \in Z_0$. Then $\Lambda$ and $\phi$ can be considered as sections of some associated vector bundles to  \fiber{Z_0}{\tQ}{M/G}. On the other hand, the $G_0$-invariance implies that
\begin{align}
& Ad_{\lambda(g_0)^{-1}} \tom = \tom  &
& Ad_{\lambda(g_0)^{-1}} \circ \Lambda \circ Ad_{g_0} = \Lambda &
& Ad_{\lambda(g_0)^{-1}} \circ \phi \circ Ad_{\lambda(g_0)} = \phi &
\label{equivariance}
\end{align}
for any $g_0 \in G_0 $, and the $(G_0\times Z_0)$-horizontality gives us that
\begin{align}
  & \Lambda(X_0) =  \phi(\lambda_* X_0)  & 
  &\forall X_0 \in \cG_0 &
  &\text{and}&
  & \tom(Z^{\tQ}_0)=  \phi(Z_0) & 
  & \forall Z_0 \in \cZ_0
\label{horizontality}
\end{align}

The ordinary $\cG$-invariant connections are recovered in formula~(\ref{decomposition}) when $\phi = \mathbb{1}$. Indeed, in this case one gets
\begin{align*}
  \pso_{|(\tq,g,h)} = Ad_{h^{-1}} \left( \tom_{|\tq} + \Lambda_{\tq}\circ\theta^G_{|g} \right) +  \theta^H_{|h}-i\theta
\end{align*}
which is to be compared to formula~(\ref{class_decomposition}). As already explained, the extra term $i \theta$ is exactly what is needed to imbed ordinary connections into the noncommutative framework.

By means of the equivariance relation~(\ref{equivariance}),  $\Lambda$ (resp. $\phi$) intertwines the representation of $G_0$ on $\cG^{\gC}$ (resp.  $\cH^{\gC}$) with the representation of $G_0$ on the algebra $M_n \simeq \vect_{\gC}(\gone, \cH)$ and by virtue of the Schur lemma, it can be decomposed in a direct sum of isomorphisms between common irreducible blocks of $\cG^{\gC}$  (resp.  $\cH^{\gC}$) and  $M_n \simeq \vect_{\gC}(\gone, \cH)$. 
Furthermore, if one requires that the connection is anti-hermitian,
one has to identify isomorphisms which correspond to complex conjuguate representations, or directly look at real representations.

This ``local'' characterization of invariant noncommutative connexions can be shown to be equivalent to the ``global'' approach. In order to do that, one needs to decomposed the degrees of freedom and rearrange them in a different way.

From (\ref{decomposition}), it is easy to  write down local expressions on $M$. A section $S : M \to \tE$ can be factorized through a local section $s=s_{\tQ}\times s_G$ on the fiber bundle $Q'=\tQ\times G $, and a section $s_{H}$ on the trivial fiber bundle $Q'\times H$. The application  $S=\Psi \circ s_H \circ s$ is represented in the following commutative diagram
\begin{align*}
  \xymatrix{
    {Q' \times H} \ar[r]^-{pr_1} \ar@{->>}[d]_-{\Psi} & {Q'=\tQ \times G} \ar@{->>}[d] \ar@/^/[l]^-{s_{H}} \\
    {\tE} \ar[r]^-{\tilde{\pi}} &   {M=M/G \times G/G_0} \ar@/_/[u]_-{s=s_{\tQ}\times s_G}  \ar@/^/[l]^-{S} }
\end{align*}
Then, the local 1-form connection is just  $S^* \omega = s^* \circ s_H^* \circ \Psi^* \tom \in \Omega(M)\otimes \Omega_\der(M_n) $. It is useful to write the section $S$ in the following way:
\begin{align*}
S : & M \longrightarrow  \tE \\
& m \longmapsto \Psi(s_{\tQ}(m),s_G(m),h(m))
\end{align*}
Then, one finally obtains
\begin{align}
S^* \omega =
Ad_{h^{-1}} \left( s_{\tQ}^* \tom +  s_{\tQ}^*\Lambda \circ s_G^*\theta^G +   s_{\tQ}^*\phi \circ Ad_{h}\circ (h^*\theta^H -i\theta)\right)
\label{local_form}
\end{align}

One can look at  passive gauge transformations which preserve symmetries of the local 1-form $S^*\omega$. We have three ways to perform such a passive gauge transformation. We can multiply on the right  $s_H$  by an element $h' \in H$, $s_{\tQ}$ by an element $z_0 \in Z_0$ or $s_G$ by an element $g_0 \in G_0$.
Then $S$ is modified in the following way
\begin{align*}
  S \longmapsto S'= \Psi \circ (s_{\tQ} \cdot z_0,s_G\cdot  g_0 ,h\cdot h' )
\end{align*}
One can report the action on $\tQ$ and $G$ to an action on $H$ by means of the equivariance of the application $\Psi$, and one has
\begin{align*}
  \Psi \circ (s_{\tQ} \cdot z_0,s_G\cdot  g_0 ,h\cdot h' ) &=  \Psi \circ (s_{\tQ},s_G ,\lambda(g_0^{-1})\cdot z_0^{-1} \cdot h \cdot h' )
\end{align*}
We will illustrate this three kind of gauge transformations in the next section.

\section{Examples}

In this section we apply the results found in the previous section to two examples. The first one is a noncommutative extension of a case extensively studied and used in the literature \cite{forgacs:80,witten:77}, that is, the spherical symmetry for $SU(2)$-gauge fields (see \cite{volkov:98} and reference therein for examples of applications). We will show that our framework generalizes in a straightforward way the results found in the ordinary case.

The second example is a purely noncommutative case. It consists to look at some symmetries on the matrix algebra. In this case, no geometry is involved.

\subsection{Spherical symmetry}
Consider the interesting example where $M=\gR \times \gR^3 \setminus \{0\}$, the first factor being parameterized by the time coordinate $t$ and  the second factor by spacial coordinates $(x,y,z)=\vec{r}$. The symmetry group is taken to be $G=SU(2)$ and it acts on $\gR^3\setminus\{0\}$ by rotation matrices\footnote{We consider the space $\gR^3\setminus\{0\}$ because we want a simple action}. Then $G_0$ is isomorphic to $U(1)$, $G/G_0$ is isomorphic to the 2-sphere $\gS^2$ and $M/G=\gR\times\gR^{+*}$. We look at a gauge theory with structure group $H=SU(2)$, and so for the noncommutative part, we  take $M_n= M_2(\mathbb{C})$.

We first treat this example in the approach developed in section~\ref{global_approach}. Notice that any principal $SU(2)$-fiber bundle over $M$ is trivial due to the fact that $M=\gR \times\gR^{+*} \times \gS^2 $, where $\gR \times\gR^{+*}$ is contractible and $\dim( \gS^2)=2$. Then  we will take $E$ in the trivial form
\begin{align*}
  E &= M\times SU(2) = M/G \times \gS^2 \times SU(2)
\end{align*}
Now, we can lift the action of $G$ on the base space $M$ to an action on the fiber bundle $E$ defining an action of $G$ on the structure group $H$. One can extend this action in a trivial way by considering the action $(g, h) \mapsto  h \ \forall g\in G, \forall h\in H$. Then the reduced theory is a $SU(2)$-gauge theory over $M/G$ which means that nothing very interesting is happening. A more complex case is to consider the following action
\begin{align*}
G\times H &\longrightarrow H\\
(g, h) &\longmapsto g\cdot h
\end{align*}
which is possible here because $G=H$.
With this action, one can see that the application $\lambda$ is an isomorphism and then one can take the reduced bundle $Q$ in such a way that $\lambda=\gone$. Then one has $Z_0= G_0=U(1)_Z$, where 
\begin{align*}
U(1)_Z :=\{\exp{(2 \epsilon T_3)}, \epsilon \in \gR\} 
\end{align*}
and $\{T_1,T_2,T_3\}$ is a basis of antihermitian generators of $\tsu(2)$ satisfying
\begin{align*}
[T_1,T_2]&= T_3 & [T_2,T_3]&= T_1 & [T_3,T_1]&= T_2 &
\end{align*}
The reduced fiber bundle $Q$ is isomorphic to $ M/G \times \{N,S\} \times U(1)$, where  $N$ and $S$ are the north and south poles of the 2-sphere $\gS^2$. Without lost of generality, we can restrict the fiber $Q$ to the point $N$, the symmetry relations coming from this $\gZ_2$ structure being just conjugation relations over complex numbers. In this case, the diagram~(\ref{diagram}) becomes
{
  \footnotesize
  \begin{align*}
    \xymatrix@R=7pt@C=3pt@M=6pt{
      U(1)_Z \ar@{^{(}->}[rrr] \ar@{^{(}->}[drr] \ar@{=}[dd] & & & \gZ_2 \ltimes U(1)_Z \ar@{>>}[rrr] \ar@{^{(}->}[drr] \ar'[d][dd] & & &\gZ_2 \ar@{^{(}->}[drr] \ar'[d][dd] & & \\  
      & & SU(2) \ar@{^{(}->}[rrr] \ar@{=}[dd] & & & \gS^2 \times SU(2) \ar@{>>}[rrr] \ar[dd] & & &\gS^2 \ar[dd] \\
      U(1)_Z \ar'[rr][rrr] \ar@{^{(}->}[drr] & & &M/G \times  \{N,S\}  \times U(1)_Z  \ar@{>>}'[rr][rrr]^-{\pi_Q} \ar@{^{(}->}[drr] \ar@{>>}'[d][dd] & & & \{N,S\}\times M/G \ar@{^{(}->}[drr] \ar@{>>}'[d][dd] & & \\
      & & SU(2) \ar[rrr] & & & M/G\times\gS^2\times SU(2) \ar@{>>}[rrr]^-{\pi} \ar@{>>}[dd] & & & M/G \times \gS^2 \ar@{>>}[dd] \\
      & & & M/G \ar@{=}'[rr][rrr] \ar@{=}[drr] & & & M/G \ar@{=}[drr] & & \\
      & & & & & M/G \ar@{=}[rrr] & & & M/G  } 
  \end{align*}
  }
Here, we have $\cL=\cM= \vect_{\gR}(T_1,T_2)$, $\cK=0$ and $\wo=\vect_{\gC}(\gone,T_3)$. It is then easy to see that $\cF\simeq \cL^{\gC}\oplus \cM^{\gC}$. Finally a $SU(2)$-invariant connection is characterized by two sections $\psi$ and $\zeta$ over $M/G$ with values in $ \vect_{\gC}(T_1,T_2)$ and a noncommutative 1-form $\mu -\eta \in (\Omega(M/G)\otimes \Omega(\wo))^1$. Here $\Omega^1(\wo)$ is simply $\wo$ because $\zo$ is 1-dimensional. If only  anti-hermitian connections are taken into account, then one can consider vector spaces over $\gR$, and $\psi$ and $\zeta$ can be interpreted as complex scalar fields (because $\cL=\cM \simeq \gC$). In this case, one has $\cC= C^{\infty}(M/G)\otimes \wo$, and $SU(\cC) =\{ e^{\chi T_3}=\cos{\frac{\chi}{2}} \, \gone +\sin{\frac{\chi}{2}} \, T_3 , \    \chi\in C^{\infty}(M/G) \} \simeq U(1)$.

In this particular example, the ``local'' approach of section~\ref{local_approach} is very well adapted because of the structure of the base space $M$. Then, we will perform the rest of its analysis using these technics. The $SU(2)$-principal bundle $E$ can be constructed from a principal $U(1)$-fiber bundle and the conjugacy class $[\lambda]=[\gone]$.

An invariant connection is given explicitely by formula (\ref{decomposition}), and for the simplicity of the analysis, we will consider only traceless  anti-hermitian connections in the following\footnote{The trace term in a connection corresponds to the term $\gone$ in the algebra $M_n=\vect_{\gC}(\gone, \cH)$, and it can be  studied   independantly to the traceless part.}. Using notations and results of section~\ref{local_approach}, we are lead to study the decomposition of the adjoint representation of $SU(2)$  in irreducible representations of $U(1)_Z$. The adjoint representation of $SU(2)$ is decomposed into the fundamental representation of $U(1)$ on $\vect_{\gR}T_3$ and the 2-dimensional  representation on $\vect_{\gR}(T_1,T_2)$, corresponding to the fundamental representation of $SO(2)$. The invariance properties~(\ref{equivariance}) implies that
\begin{align*}
  \Lambda(T_1) &=\Lambda_1 T_1 + \Lambda_2 T_2 &
  &&
  \phi(T_1) &=\phi_1 T_1 + \phi_2 T_2 \\
  \Lambda(T_2) &= -\Lambda_2 T_1 + \Lambda_1 T_2 &
  &&
  \phi(T_2) &= -\phi_2 T_1 + \phi_1 T_2
\end{align*}  
 and using (\ref{equivariance}) and (\ref{horizontality})
 \begin{align*}
& \Lambda(T_3)=\phi(T_3)= \eta T_3
\end{align*}
where $\eta$ is a function over $M/G$.
Now let us write the local expression~(\ref{local_form}) of the connection 1-form by considering two different useful gauges. First let us introduce what we call the ``singular'' gauge in which we take the constant section 
\begin{align*}
s_H : Q' &\to Q' \times H \\
q' &\mapsto (q',e)
\end{align*}
We will choose the usual spherical coordinates $(\vartheta,\varphi)$ for the local system of coordinate on $\gS^2$. Then we consider the natural local section 
\begin{align*}
  s_{G} : \gS^2 &\to SU(2) \\
  (\vartheta, \varphi) &\mapsto g=e^{\varphi T_3}e^{\vartheta T_2}
\end{align*}
(we have in mind the Euler parametrisation of $SU(2)$, where $\vartheta$ and $\varphi$ are two of the three Euler angles).
Then a straightforward computation gives 
  \begin{multline}
    S^* \omega = a T_3 + (\Lambda_1 T_1 + \Lambda_2 T_2) \d \vartheta + (\Lambda_1 T_2 - \Lambda_2 T_1) \sin\vartheta \d \varphi + \eta  T_3 \cos\vartheta \d \varphi \\
      - \left[  (\phi_1 T_1 +\phi_2 T_2) \theta^1  + (\phi_1 T_2- \phi_2 T_1) \theta^2 + \eta T_3\theta^3 \right]  
  \label{singular_form}
  \end{multline}
where $a=a_r \d r + a_t \d t \in \Omega^{1}(\gR\times \gR^{+*})$, and $i \theta= T_a \theta^a$. This reduced 1-form connection  generalizes in an obvious way the so-called Witten's anzatz~\cite{witten:77} which is recovered by setting $\phi_1=\eta=1$ and $\phi_2=0$ (i.e. $\phi = \gone$). One can note that the monopole term (corresponding to the local 1-form $\cos\vartheta \d \varphi$) is no more constant and is  now factorized by a function $\eta$. Singularities happening in~(\ref{singular_form}) are due to the fact that we try to extend the system of spherical coordinates globally on $\gS^2$. This extension is not possible in this gauge and it is why it is called the ``singular'' gauge. However one can introduce an other gauge in which the extension of the local 1-form to a global one is possible. This is the ``regular'', or ``radial'' gauge, defined by the following section
\begin{align}
S : M &\longrightarrow \tE \\
(r,\vartheta,\varphi) &\longmapsto \Psi (s_{\tQ}(r), e^{\varphi T_3}e^{\vartheta T_2},e^{-\vartheta T_2}e^{-\varphi T_3})
\label{section0}
\end{align}
It can be obtained from the  ``singular'' gauge by a (passive) gauge transformation which consist to multiply  the section $s_H$ by the element $h'= e^{-\vartheta T_2}e^{-\varphi T_3} \in H$. Then applying formula~(\ref{local_form}) leads to the following expression
\begin{equation}
  \begin{split}
{S}^* \omega = a T_r 
   & + \Lambda_1 [T_r, \d T_r] - \Lambda_2 \d T_r \\
    &  -\phi_1 [T_r, \hd T_r] +\phi_2 \hd T_r  - \eta T_r \theta^r  
  \end{split}\label{radial_gauge}
\end{equation}
where 
\begin{align*}
  T_r&= \sin\vartheta \cos\varphi \ T_1 + \sin\vartheta \sin\varphi \ T_2+  \cos\vartheta \ T_3 \\
  \theta^r &= \sin\vartheta \cos\varphi \ \theta^1 + \sin\vartheta \sin\varphi \ \theta^2+ \cos\vartheta \ \theta^3
\end{align*}
and $\hd= \d+ \d'$ is the noncommutative differential  introduced in section~\ref{ncdc}. The absence of singularity is clearly due to the fact that the spherical angles $(\vartheta, \varphi)$ do not appear explicitly, and that everything can be expressed in term of the unique generator $T_r$. To illustrate the fact that we have extended the local 1-form to a global one, we will give some more explicit formulas  using euclidian coordinates. Let us introduce the notation  $S^*\omega= a + A^a_i T_a \d x^i -  \phi^a_b T_a \theta^b $. Then formula~(\ref{radial_gauge}) gives us
\begin{align*}
A_i^a&=\frac{\Re(\psi -i \phi)}{r} \ P^a_i \ + \ \frac{\Im(\psi -i \phi)}{r} \ g^{ab}\epsilon_{ibc} \hn^c\\
\phi_b^a&= \Re(\phi) \ P^a_b \ +\  \Im(\phi) \ g^{ac}\epsilon_{bcd}\hn^d \  + \ \eta\ \hn^a \hn_b
\end{align*}
with  $\hn^a=\frac{x^a}{r}$, $P^a_b=\delta^a_i-\hn^a\hn_i$, $g^{ab}$ the euclidian metric and $\epsilon_{abc}$ the totally antisymmetric tensor such that $\epsilon_{123}=1$. We have introduced the useful  notations  $\psi=-\Lambda_2+ i \Lambda_1$ and $\phi=\phi_1 + i \phi_2$.

Finally, we would like to show how the two other passive gauge transformations (on $s_G$ and $s_{\tQ}$ mentioned at the end of section~\ref{local_approach}) can be performed. They will correspond  to  a $U(1)$ residual symmetry. In term of the two complex scalar fields $\psi$ and $\phi$, equation~(\ref{radial_gauge}) becomes
  \begin{multline*}
    S^* \omega = a T_r 
    +  \Re(\psi -i \phi) \d T_r  + \Im(\psi -i \phi) [T_r, \d T_r] \\
    + \Re(-i \phi) \d' T_r + \Im(-i \phi) [T_r,\d' T_r] - \eta T_r \theta^r
  \end{multline*}
The $U(1)$ passive gauge transformations correspond to the transformations on $s_G$ and $s_{\tQ}$ given by
\begin{align*}
  s_G &\leadsto s_G\cdot e^{\chi_0 T_3}\\
  s_{\tQ} &\leadsto s_{\tQ}\cdot e^{\chi_1 T_3}
\end{align*}
where $\chi_0(r,t)$ and $\chi_1(r,t)$ are two arbitrary functions of $r$ and $t$. This leads to  transformations on the fields $\psi$ and $a$: 
\begin{equation}
\begin{split}
\psi &\leadsto e^{i(\chi_1+\chi_0)} \psi \\
a &\leadsto a - \eta \d (\chi_1+\chi_0)
\end{split}
\label{gauge-transformation}
\end{equation}
One can remark that the  scalar fields $\phi$ and $\eta$ remain unchanged under this passive gauge transformations. One can also note the similarity with usual abelian gauge transformations.

It is possible to consider a ``true'' symmetric gauge transformation (in the sens of noncommutative geometry, see section~\ref{ncdc}). In order to do that, one has  to redefine the complex scalar field $\phi$ to $\phi = 1- \phi'$. Then a ``true'' symmetric gauge transformation parametrised by an element $e^{\chi T_3} \in SU(\cC)$, where $\chi$ is a function over $M/G= \gR\times\gR^{+*}$, leads to the transformations
\begin{equation}
\begin{split}
\phi'  &\leadsto e^{-i\chi} \phi' \\
\psi &\leadsto e^{-i \chi} \psi \\
a &\leadsto a + \d \chi
\end{split}
\end{equation}
Note that these transformations are much more similar to ordinary $U(1)$ gauge transformations.

\subsection{A purely noncommutative example}
We present in this section a purely noncommutative case in the sense that we consider a situation in which  the base space is a point. Then one has $G=G_0$, and $\lambda $ is an homomorphism from $G$ to $H$.  In this case, the noncommutative algebra is simply the algebra of matrices $M_n$, equipped with  the noncommutative calculus exposed  in~\cite{dubois-violette:90} and summarized in section~\ref{ncdc}. In this situation, the problem reduces to characterize  $G$-invariant noncommutative connections in $M_n$. For simplicity, as before, we will restrict in this section to traceless connections. 

The general procedure to follow is to first study the representation $\lambda$ of $G$ in $M_n$, and then determine how the representations $Ad^{H}\circ\lambda$ split into irreducible representations of $G$. These irreducible representations give rise to the degrees of freedom of the invariant noncommutative connections: one scalar field for each intertwiner of equivalent irreducible representations.

We want to illustrate this in the particular case $G=SU(2)$. It is well known that representations of $SU(2)$ on $M_n$ are parameterized by partitions of $n$ \cite{dubois-violette:japan99}. For instance, for $\tA=M_3(\gC)$, the representations are labeled by the partitions  ``1+1+1'', ``2+1'' and ``3''  of $3$. They are  described in the following way:
\begin{itemize}
\item The ``1+1+1'' representation corresponds to the sum of $3$ copies of the trivial representation.  The representation $Ad^{H} \circ \lambda$ is decomposed in a sum of $8$ copies of the trivial  representation of $SU(2)$. This gives rise to 64 scalar fields. This case is not interesting because the group $SU(2)$ does not act on $M_3(\gC)$.
\item the ``2+1'' representation corresponds to  a reducible representation of $SU(2)$ which is the sum of the fundamental representation and the trivial one. The representation $Ad^{H} \circ \lambda$ is decomposed into irreducible representations of $SU(2)$ of dimensions 3, 2, 2 and 1 (6 scalar fields).
\item the ``3'' representation corresponds to the irreducible representation of dimension 3 of $SU(2)$. In this case, the representation   $Ad^{H} \circ \lambda$ can be decomposed into irreducible representations of $SU(2)$ of dimensions 3 and 5 (2 scalar fields).
\end{itemize}
We will not perform the analysis any further because our goal is just to characterize the degrees of freedom of a traceless noncommutative connections using the methods exposed in this paper.

\subsection*{Acknowledgments}
The authors would like to thank M. Dubois-Violette, Y. Georgelin,  M. Maceda and J. Madore, for interesting discussions about various parts of this work.

\bibliographystyle{utphys}
\bibliography{biblio_articles}
\end{document}